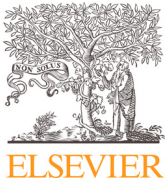
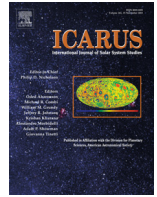

# Anelastic spherical dynamos with radially variable electrical conductivity

W. Dietrich [a,b,*], C.A. Jones [a]

[a] *Department of Applied Mathematics, University of Leeds, Leeds LS2 9JT, United Kingdom*
[b] *Max Planck Institute for Solar System Research, Justus-von-Liebig-Weg 3, 37077 Goettingen, Germany*

**A B S T R A C T**

A series of numerical simulations of the dynamo process operating inside gas giant planets has been performed. We use an anelastic, fully nonlinear, three-dimensional, benchmarked MHD code to evolve the flow, entropy and magnetic field. Our models take into account the varying electrical conductivity, high in the ionised metallic hydrogen region, low in the molecular outer region. Our suite of electrical conductivity profiles ranges from Jupiter-like, where the outer hydrodynamic region is quite thin, to Saturn-like, where there is a thick non-conducting shell. The rapid rotation leads to the formation of two distinct dynamical regimes which are separated by a magnetic tangent cylinder - mTC. Outside the mTC there are strong zonal flows, where Reynolds stress balances turbulent viscosity, but inside the mTC Lorentz force reduces the zonal flow. The dynamic interaction between both regions induces meridional circulation. We find a rich diversity of magnetic field morphologies. There are Jupiter-like steady dipolar fields, and a belt of quadrupolar dominated dynamos spanning the range of models between Jupiter-like and Saturn-like conductivity profiles. This diversity may be linked to the appearance of reversed sign helicity in the metallic regions of our dynamos. With Saturn-like conductivity profiles we find models with dipolar magnetic fields, whose axisymmetric components resemble those of Saturn, and which oscillate on a very long time-scale. However, the non-axisymmetric field components of our models are at least ten times larger than those of Saturn, possibly due to the absence of any stably stratified layer.

## 1. Introduction

The magnetic fields and zonal wind structure of Jupiter and Saturn can be modeled using the anelastic MHD spherical dynamo equations. A key feature distinguishing Saturn from Jupiter is the depth where the transition between molecular hydrogen and its high-pressure metallic phase occurs (Lorenzen et al., 2011). Jupiter models, such as Jones (2014) or Duarte et al. (2013), which successfully reproduce the magnetic field dipolarity and dipole tilt, have a shallow hydrodynamic layer and a thick dynamo region deeper down. We therefore extend and generalise these studies by using the thickness of the deeper dynamo zone as a parameter, considering the resulting differential rotation and magnetic fields. In particular, we discuss the implications for Saturn. Gas giant planets have deep atmospheres with a radial outward decay of static density, pressure and temperature, probably overlying a relatively small rocky core. This makes a compressible approach, such as the anelastic approximation, a useful basis on which to build models.

The main challenge of numerical MHD models covering the global atmospheric dynamics and magnetic field induction process of gas giants stems from the enormous range of time and length scales characteristic of a rapidly rotating, planet-sized spherical fluid body. Additionally, electric currents and their associated magnetic forces constitute another level of complexity. The nonlinear interaction and relative importance of the governing forces (buoyancy, Coriolis, Lorentz, dissipation) leads to characteristic phenomena such as predominantly columnar convective flows, deep-reaching zonal wind systems, and dynamo-generated magnetic fields. The ratios of those forces are typically quantified by a set of nondimensional numbers, e.g. the Ekman, Rayleigh, hydrodynamic and magnetic Prandtl number. For some of these, e.g the Ekman number, the natural value is far beyond that possible in numerical models. We must therefore use enhanced diffusivities and surface heat flux, hoping that the small scales they eliminate are not so important in determining the larger scale flows and fields we are most interested in.

* Corresponding author at: Max Planck Institute for Solar System Research, Justus-von-Liebig-Weg 3, 37077 Goettingen, Germany.
  *E-mail address:* dietrichw@mps.mpg.de (W. Dietrich).

*1.1. Zonal flows*

Both Saturn and Jupiter feature a prograde jet at the equator, which is broader and more energetic on Saturn, and multiple alternating bands at higher latitudes (Sanchez-Lavega et al., 2000; Porco et al., 2003). One possible cause for the surface zonal flow structure are rotationally organised convective motions filling the whole planetary atmosphere and generating geostrophic differential rotation. There have been plenty of numerical and theoretical studies regarding the emergence of differential rotation in rotating convection. For hydrodynamic models, convection emerges in accord with the Taylor–Proudman theorem as columnar structures aligned with the rotation axis (Busse, 1970). The columns exhibit a consistent tilt leading to Reynolds stress, which pumps zonal angular momentum outwards, e.g. Busse (1976); Christensen (2001); Aubert (2005). This leads to a run-away growth of differential rotation ultimately balanced by the tiny viscosity and hence giving enormous jet amplitudes.

More recently, such zonal flows have been successfully modelled in anelastic systems as well (e.g. Jones and Kuzanyan, 2009; Gastine and Wicht, 2012). Those models are usually strongly geostrophic, and harbour a broad prograde jet close the equator and a retrograde one at greater depth. Low Ekman number global 3D models can give Jupiter-like zonal flows, but only if the model is restricted to the non-metallic region, and crucially uses a stress-free lower boundary (Heimpel et al., 2005).

The magnetic field created in the deep interior can attenuate the zonal flow and alter its pattern. Earlier models of giant planet atmospheres including the magnetic field relied on the simpler Boussinesq approach (Heimpel et al., 2005; Heimpel and Aurnou, 2007; Gómez-Pérez et al., 2010). More recent studies used a polytropic perfect gas, or an interior state model covering the strong radial decline of density, temperature and pressure in the framework of the anelastic approximation (Gastine et al., 2012; Duarte et al., 2013; Jones, 2014). Those models indicated that when the dynamo is active, the main force balance is altered from a geostrophic to a magnetostrophic regime. Then the run-away growth of differential rotation due to Reynolds stresses is stopped in the metallic hydrogen region by the influence of Maxwell stresses. Such models successfully reproduced the magnetic field morphology of Jupiter and the equatorial prograde jet (Jones, 2014). However, the high latitude alternating structures in giant planets are suppressed by the Maxwell stresses, so it is more likely they are a surface effect not related to the deep interior. Ultimately, the depth of the surface zonal flows may be observationally constrained with gravity measurements by the Juno mission (Hubbard, 1982; Kaspi et al., 2010; Zhang et al., 2015).

*1.2. Magnetic fields, parity and classification*

Jupiter's magnetic field is rather Earth-like in terms of dipolarity and mean dipole tilt (Connerney, 2007). However, Saturn's field is peculiar as it seems entirely axisymmetric (e.g. Smith et al., 1980; Cao et al., 2011). Stevenson (1982) suggested that zonal flows in a stably stratified layer could make the surface magnetic field axisymmetric even if the field generated by the dynamo is significantly non-axisymmetric. Such a scenario was further explored in numerical models by Christensen and Wicht (2008); Stanley (2010); Stanley and Bloxham (2016). However, these models ignored the other main characteristic of Saturn's atmospheric interior, the variable radial conductivity and the combination of a deep dynamo and an outer hydrodynamic shell. In this study, we exclude the uncertain stably stratified layer that may occur in Saturn, and study the induction of magnetic fields when the electrical conductivity is a function of radius. Then, from the modeling point of view, the essential difference between Jupiter and Saturn is the depth of the conductivity drop-off level.

The magnetic fields of dynamo models are usually classified by the percentage contribution of the dipole mode (Duarte et al., 2013). It turned out that dipole dominated fields are rather hard to isolate in the most advanced Jupiter models (Jones, 2014). They seem to coexist with other leading order field symmetries in close proximity in the parameter space explored so far. It is also known that strong anelasticity and the use of smaller (more realistic) hydrodynamic Prandtl numbers yield a rich zoology of dynamos (Simitev and Busse, 2005; Christensen and Aubert, 2006; Duarte et al., 2013). Though the global magnetic fields of the Sun, the Earth, Jupiter and Saturn are predominantly dipolar, the ice giants and also Mercury are found to harbour strong non-dipolar contributions, including a significant quadrupolar component.

The process of selecting a leading order equatorial symmetry is not well understood. In terms of the simpler kinematic theory, Roberts (1972) and Proctor (1977) showed that a combination of $\alpha$-effect, shear and meridional circulation controls the time-dependence and leading order field parity. Their results suggest that, steady dipolar and quadrupolar dynamos are typically excited at similar parameter values in $\alpha^2$-dynamos. The presence of strong shear gives preference to dipolar waves of $\alpha\Omega$-type (Parker, 1955). If there exists additionally a significant meridional circulation, either quadrupolar or dipolar steady solutions are selected by the sign of the product of the $\alpha$-effect and $\Omega$ (Proctor, 1977). Later on, Sreenivasan and Jones (2011) proposed that the magnetic field itself enhances the efficiency of the induction process by increasing the kinetic helicity, leading to a dipolar preference over all.

Dynamo generated global magnetic fields in natural objects, such as planets or stars, can vary substantially in field symmetry and time-dependence. Magnetic fields with leading quadrupolar (equatorially symmetric) parity have been observed in stellar dynamos, e.g. for Bp-type stars (Thompson and Landstreet, 1985; Kochukhov, 2006). These stars are substantially more massive than the Sun ($M > 1.5 M_\odot$) hence they are convectively unstable in the core and stably stratified in the radiative outer region, which is known to develop differential rotation (Triana et al., 2015). Further the quadrupolar magnetic moment of the Sun can be substantial at periods during the solar cycle, and was suggested to dominate over the axial and equatorial dipole components at the time of grand solar minima (Knobloch et al., 1998; Beer et al., 1998; DeRosa et al., 2011). The magnetic fields of the ice giants are best described with a strong contribution from the equatorial dipole mode (Holme and Bloxham, 1996; Stanley and Bloxham, 2004). Hence they have substantial equatorially symmetric and non-axisymmetric components. Finally, the dynamos in low-mass dwarfs generate an azimuthal magnetic structure harbouring a wide variability of axisymmetric and non-axisymmetric modes (Donati, 2011).

*1.3. Radially variable electrical conductivity*

In Jupiter and Saturn the transition from metallic to molecular hydrogen leads to a steep decrease in the electrical conductivity (Chabrier et al., 1992; French et al., 2012). The conductivity profile then follows an exponential decay in the metallic hydrogen region deeper inside accompanied by a super-exponential decay outside the metallic hydrogen region (Jones, 2014). This implies an active magnetic field generation process deep inside the planet where hydrogen exhibits metallic electrical conductivity. The drop-off radius separating the hydrodynamic outer shell from the magnetic interior shell is closer to the surface for massive Jupiter (at 90% of the planetary radius), but is much deeper for Saturn (at 65%-70%). From the modeling perspective, this leads to an inner conducting shell where the magnetic field dominates the dynam-ics surrounded by an outer hydrodynamic shell where the strong Coriolis force reigns. For Saturn, it has been suggested that a stably stratified region is formed underneath the conductivity drop-off, as He might become immiscible (Stevenson and Salpeter, 1977). However, more recently it was shown that the He-droplets sink down towards the solid core and may not build up a stratifying compositional gradient between the deep-seated dynamo and non-metallic shallower regions (Püstow et al., 2016). Hence we do not include any stable stratification in the models and focus on the effect of a radially varying electrical conductivity.

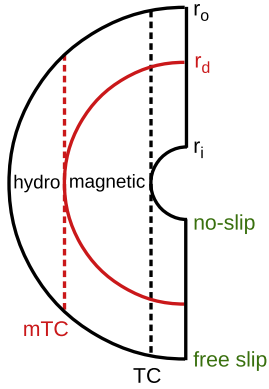

**Fig. 1.** Schematic model setup denoting the inner, no-slip and outer, free slip wall. The imaginary cylindrical boundary attached to the inner boundary is termed the tangent cylinder (TC). The red line denotes the radial position of the conductivity drop-off ($r_d$) separating an inner magnetic shell with high electrical conductivity harbouring the dynamo process from an outer hydrodynamic one. The red dashed line gives the position of the cylindrical magnetic tangent cylinder (mTC). (For interpretation of the references to colour in this figure legend, the reader is referred to the web version of this article.)

A non-constant electrical conductivity has been investigated in Boussinesq systems (Gómez-Pérez et al., 2010; Heimpel and Gómez Pérez, 2011) and recently in anelastic models (Duarte et al., 2013; Jones, 2014; Gastine et al., 2014). These studies focused on Jupiter, hence the effect of a deeper metallicity region has not yet been fully explored. Our models are based upon the study of Jones (2014), but investigate a deeper conductivity drop-off, more applicable to Saturn-sized planets, and cover a broader range in model parameters. Gómez-Pérez et al. (2010) and Heimpel and Gómez Pérez (2011) found that models with thick non-conducting regions have much stronger zonal flows than those which are magnetic almost everywhere. An important issue is at what radius the magnetic field first becomes dynamically important. A spherical boundary termed the 'planetary tachocline' at a depth of $r_d$ was suggested to separate the hydrodynamic outer from the magnetic inner region (Gómez-Pérez et al., 2010; Heimpel and Gómez Pérez, 2011). We give a precise definition of $r_d$ below, but it coincides roughly with the transition from the non-magnetic H/He region to the metallic hydrogen region. Our results suggest the boundary between the hydrodynamic and the magnetically controlled region has cylindrical geometry, a 'magnetic tangent cylinder' - mTC (see Fig. 1) due to the strong rotational influence. This virtual cylinder is attached to $r_d$ at the equator.

This indicates that surface zonal flows outside the mTC are deep reaching, geostrophic differential rotation systems, whereas the alternating jet structures observed at higher latitudes on Saturn and Jupiter are rather shallow phenomena (Jones, 2014).

Varying $r_d$ effectively changes the geometry of the dynamo region. This shares some similarities with models featuring constant conductivity, but where the aspect ratio is changed. Goudard and Dormy (2008) found in a Boussinesq model that when the dynamo aspect ratio exceeds $r_d/r_i = 0.65$, the previously preferred steady dipole solutions are replaced by oscillating dipolar or quadrupolar solutions. Interestingly, the dynamos are reported to jump between the dipolar and quadrupolar branch over their temporal evolution (Goudard and Dormy, 2008).

We start off the paper by discussing the implementation of a radially variable electrical conductivity in Section 2. There follows the introduction of the MHD model and the computational aspects (Section 3). Section 4 then provides a detailed description of the results subdivided into an analysis of zonal flows and their maintenance (Section 4.1), the kinetic helicity (Section 4.2), a classification of the emerging dynamo solutions in terms of butterfly diagrams (Section 4.3) and the 'magnetic trigram' (Section 4.4). The results are compared to Saturn in Section 4.5, before the paper concludes in Section 5.

## 2. Variable electrical conductivity and magnetic Reynolds number

To model the radially varying magnetic diffusivity ($\lambda^{-1} = \mu_0 \sigma$, with $\sigma$ the electrical conductivity), a hyperbolic fitting formula (Jones, 2014)

$$\lambda(r) = \exp\left(u + \sqrt{u^2 + v}\right), \tag{1}$$

where $u$ and $v$ are

$$u = \frac{1}{2}[(g_1 + g_2)r - g_2 - g_4] \tag{2}$$

$$v = (g_1 r - g_2)(g_3 r - g_4) - g_5, \tag{3}$$

$r$ being radius in metres, was used. With the values $g_1 = -4.279 \cdot 10^{-6}$, $g_2 = 274$, $g_3 = -2.55 \cdot 10^{-8}$, $g_4 = 1.801$, $g_5 = 20.28$ this gives a close fit to the electrical conductivity of the Jupiter model of French et al. (2012). For our models $g_2$ and $g_5$ are changed to alter $r_d$, the conductivity drop-off. The others are kept constant. The five different profiles with $g_2 = 274, 240, 210, 195$ and $180$ are plotted in Fig. 2, top panel. Additionally, $g_5 = 20.28$ for all cases, with the exception of the ($g_2 = 180$)-model, where $g_5 = 10.0$ to keep the diffusivity in the dynamo region comparable to the other profiles. The model with $g_2 = 274$ (green profile) closely resembles the Jupiter-profile used in Jones (2014). Whereas for Saturn the phase transition from molecular to metallic hydrogen is assumed at the 1 Mbar-level at $0.67 r_S$ (Nettelmann et al., 2013), so the orange profile is the closest to Saturn. For numerical reasons, we set the gradient of $\lambda$ to $d\lambda/dr = 10^6$ if the formula value is larger, assuming that this gives a diffusivity sufficiently large to yield a current-free region for $r > r_d$. Finally, all conductivity profiles are normalised to their respective mid-depth values ($\lambda_m$).

To roughly indicate the top of the dynamo region, located at the conductivity drop-off $r_d$, we quantify the induction of magnetic field by estimating the magnetic Reynolds number, $Rm$. The diffusivity is non-constant hence the magnetic diffusion consists of two parts:

$$\nabla \times \lambda(r) \nabla \times \boldsymbol{B} = \lambda(r) \nabla \times \nabla \times \boldsymbol{B} + \frac{\lambda}{d_\lambda} \hat{\boldsymbol{e}}_r \times \nabla \times \boldsymbol{B}, \tag{4}$$

where $[d_\lambda]^{-1} = 1/\lambda \, d\lambda/dr$. In the transition region between the magnetic interior and the hydrodynamic outer shell, the second term dominates the magnetic diffusion and hence we use

$$Rm^{\star}(r) = U_{eq}(r) \, d_\lambda / \lambda(r) = U_{eq}(r) \left[\frac{d\lambda}{dr}\right]^{-1},$$

where $U_{eq}(r)$ is the rms flow strength in the equatorial plane for the models in group 2 in Table 1. Because the zonal flows are strongly damped in the magnetic region, they do not significantly affect $U_{eq}$ there. Fig. 2, bottom panel (solid lines) shows $Rm^*$ as a function of $r$. It can be seen that $Rm^*$ roughly resembles the conductivity profile (top panel); values of $Rm^* < 10^{-2}$ are effectively zero in our work. Following Heimpel and Gómez Pérez (2011), the upper border of the dynamo region $r_d$ for each profile is defined roughly where $Rm^*$ crosses unity (Fig. 2, bottom panel). Note that above $r_d$ the magnetic field is approximately a potential field as there are no significant electrical currents there. We also show in Fig. 2 (bottom panel) the dashed curves derived from setting $Rm = U_{eq}(r)D/\lambda(r)$, which would be obtained if the second term in (4) were ignored. The profile of $Rm$ is significantly larger, showing that the additional dissipation due to the variation of $\lambda$ is significant.

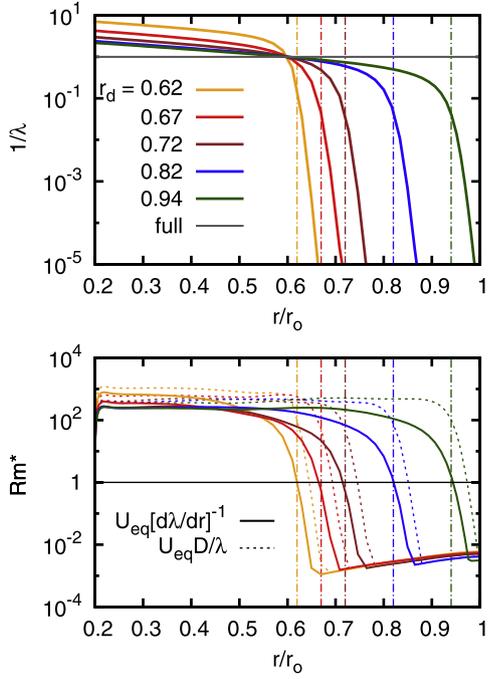

**Fig. 2.** Top: Radial decay of normalised inverse magnetic diffusivity. All curves are scaled version of the profile used in Jones (2014), where the drop-off radius $r_d$ is varied. Bottom: Magnetic Reynolds number $Rm^* = U_{eq}[d\lambda/dr]^{-1}$ (solid) compared to the generic definition $Rm = U_{eq}D/\lambda$ (dashed) along radius. The vertical, dashed lines denote the $r_d$ values defined by $Rm^* = 1$ and hence the upper border of the magnetically active region.

## 3. Models and methods

We model the unstably stratified and partially electrically conducting atmosphere of a gas giant as a rapidly rotating and vigorously convecting spherical shell. We take into account the radial variation of adiabatic temperature, density, pressure and electrical conductivity according to an interior state model originally developed for Jupiter (French et al., 2012). Note, interior state models dedicated to Saturn are available, but they do not yet cover the electrical transport properties (Nettelmann et al., 2013). It is unlikely that there is great sensitivity to the profiles of temperature, density and pressure, so using scaled Jupiter values should be adequate. As discussed in the previous study by Jones (2014), the governing equations for the conservation of mass, momentum, thermal energy and magnetic field are

$$0 = \nabla \cdot (\bar{\rho}\boldsymbol{u}), \quad (5)$$

$$\frac{\partial \boldsymbol{u}}{\partial t} + \boldsymbol{u} \cdot \nabla \boldsymbol{u} = -\frac{Pm}{E}\nabla \hat{p} - \frac{2Pm}{E}\hat{\boldsymbol{e}}_z \times \boldsymbol{u} + Pm\boldsymbol{F}_\nu$$

$$- \frac{RaPm^2}{Pr}\frac{d\bar{T}}{dr}S\hat{\boldsymbol{e}}_r + \frac{Pm}{E\bar{\rho}}(\nabla \times \boldsymbol{B}) \times \boldsymbol{B}, \quad (6)$$

$$\frac{\partial S}{\partial t} + \boldsymbol{u} \cdot \nabla S = \frac{Pm}{Pr}\frac{1}{\bar{\rho}\bar{T}}\nabla \cdot \bar{\rho}\bar{T}\nabla S + \frac{1}{Pm}\frac{Pr}{Ra\bar{T}}Q_\nu$$

$$+ \frac{Pm}{Pr}H + \frac{Pr}{RaPm\bar{T}}\frac{\lambda(r)}{E\bar{\rho}}(\nabla \times \boldsymbol{B})^2, \quad (7)$$

$$\frac{\partial \boldsymbol{B}}{\partial t} = \nabla \times (\boldsymbol{u} \times \boldsymbol{B}) - \nabla \times (\lambda(r)\nabla \times \boldsymbol{B}). \quad (8)$$

Here $\boldsymbol{u}$ is the flow, $\boldsymbol{B}$ the magnetic field, $S$ the specific entropy, $\boldsymbol{F}_\nu$ the viscous force, $\hat{p} = p/\bar{\rho}$ the modified pressure, $\hat{\boldsymbol{e}}_z$ the axis of rotation, $\hat{\boldsymbol{e}}_r$ the radial unit vector, $H$ is an entropy source, $Q_\nu$ represents viscous heating, $\bar{\rho}$ and $\bar{T}$ are density and temperature of the background. Further

$$\boldsymbol{F}_\nu = \frac{1}{\bar{\rho}}\left[\frac{\partial}{\partial x_j}\bar{\rho}\left(\frac{\partial u_i}{\partial x_j} + \frac{\partial u_j}{\partial x_i}\right) - \frac{2}{3}\frac{\partial}{\partial x_i}\bar{\rho}\frac{\partial u_j}{\partial x_j}\right], \quad (9)$$

$$Q_\nu = \sigma_{ij}\frac{\partial u_i}{\partial x_j}, \quad (10)$$

$$\sigma_{ij} = \nu\bar{\rho}\left(\frac{\partial u_i}{\partial x_j} + \frac{\partial u_j}{\partial x_i} - \frac{2}{3}\delta_{ij}\nabla \cdot \boldsymbol{u}\right), \quad (11)$$

where $\sigma_{ij}$ is the stress tensor. Background temperature, density and magnetic diffusivity are normalised by their respective mid-depth values ($T_m$, $\rho_m$, $\lambda_m$). The non-dimensional parameters emerging by rescaling length by the shell thickness $D = r_o - r_i$, time by the magnetic diffusion time scale $\tau_\lambda = D^2/\lambda_m$, and the magnetic field by $\sqrt{\Omega\rho_m\mu_0\lambda_m}$, are the Ekman number $E$, the hydrodynamic Prandtl number $Pr$, the Rayleigh number $Ra$ and the magnetic Prandtl number $Pm$ according to

$$E = \frac{\nu}{\Omega D^2}, \quad Pr = \frac{\nu}{\kappa}, \quad Ra = \frac{T_m D^3 q_o}{\rho_o T_o \kappa^2 \nu}, \quad Pm = \frac{\nu}{\lambda_m}, \quad (12)$$

where $\kappa$ is entropy diffusivity, $\nu$ kinematic viscosity and $\Omega$ the rotation rate. Here $\rho_o$ and $T_o$ are density and temperature at the outer boundary ($r = r_o$). Note that the entropy scale and hence the Rayleigh number in the present study are based on the outer boundary heat flux density $q_o$, rather than an imposed entropy contrast $\Delta S$ as used in Jones (2014). Then $q_o$, which is constant over the planet's surface, is determined by the contributions of internal ($H$) and bottom heat source density ($q_i$)

$$r_o^2 q_o = \frac{r_o^3 - r_i^3}{3}H + r_i^2 q_i, \quad (13)$$

which can be expressed using the aspect ratio $\beta = r_i/r_o$ to give

$$q_o = \frac{1 - \beta^3}{3(1 - \beta)}HD + \beta^2 q_i. \quad (14)$$

For giant planets with relatively small cores, the heat flux emerging from the core is likely to be small compared to the heat flux in the H/He region, so for the bulk of the models we ignore any heat flux at the inner boundary ($q_i = 0$), and the convection is powered exclusively by an internal entropy source $H$ due to cooling of the planet. However, for comparison we also performed a few models with bottom driving ($H = 0$, $q_i = q_o/\beta^2$). Further, the aspect ratio is increased to a Saturn-like value, $\beta = 0.2$. The containing walls are impenetrable and no-slip at the inner boundary and free-slip at the outer boundary (see also Fig. 1).

We numerically integrate the system of equations (Eqs. 5–8) by using the Leeds Anelastic Spherical Dynamo Code (ALSD), an MPI-parallelised pseudo-spectral code benchmarked against several other comparable numerical implementations (Jones et al., 2011). The radial resolution is $N_r = 160$ and spectral resolution is truncated at maximum degree and order $N_\ell = N_m = 128$ yielding azimuthal and latitudinal resolution of $N_\phi = 384$ and $N_\theta = 192$, respectively. Table 1 contains a selection of our dynamo simulations where the individual models are organised in groups, each group aiming to identify one parameter dependence. For group 1, a smaller $Pr = 0.15$ is combined with a smaller Rayleigh number and $r_d$ is varied. Groups 2 and 3 repeat the numerical experiment with slightly higher $Pr$ and different $Ra$-values. Group 4 tests the $Pr$-dependence, whereas groups 5 and 6 investigate the effect of increasing $Ra$ for fixed $r_d$. Group 7 checks models with higher magnetic Reynolds number by increasing $Pm$, whereas group 8 shows the influence of bottom driven convection. The cases 9–12 test the robustness of the regimes with various combinations of $Pr$, $Ra$ and $r_d$. Case 13 in Table 1 represents the Jupiter-like model from Jones (2014). At the chosen parameter regime the numerical simulations are extremely resource demanding. A single model requires one or two weeks of run-time, when parallelised over 512 cores, to integrate past the initial transients and time-average over a significant fraction of the magnetic diffusion time. The total computational demand for the 34 runs in the table sums up to $3 \cdot 10^6$ CPU hrs and is provided by supercomputing facilities. Simulations were initialised from previously obtained solutions, e.g. from the strong dipolar Jupiter-like model of Jones (2014). However, we also performed a few test runs initialised from a random seed field, where the Lorentz force is initially marginal.

**Table 1**

Table of runs. The Ekman number is fixed to $E = 5 \cdot 10^{-5}$, the magnetic Prandtl number to $Pm = 3$ apart from the cases where $Pm = 5$ is used ([5], group 7). D, Q and NA are the time averaged relative magnetic energy at the surface in equatorial-antisymmetric/axisymmetric, equatorial-symmetric/axisymmetric and non-axisymmetric magnetic energy at the surface, respectively. † indicates bottom heated models. ♯ denotes the Jupiter-model by Jones (2014), which was performed with a lower Ekman number $E = 2.5 \cdot 10^{-5}$, $\beta = 0.0963$ and internal heating. The column 'type' assigns each dynamo solution to the seven types given in Fig. 6. The bold-faced ones are those shown in the figure.

| o. | Ra | Pr | $r_d$ | D | Q | NA | Type | Symbol |
|---|---|---|---|---|---|---|---|---|
| 1.1 | $6 \cdot 10^6$ | 0.15 | 0.67 | 0.914 | 0.002 | 0.085 | III | ♦ |
| 1.2 | $6 \cdot 10^6$ | 0.15 | 0.72 | 0.934 | 0.014 | 0.005 | III | ■ |
| 1.3 | $6 \cdot 10^6$ | 0.15 | 0.82 | 0.004 | 0.813 | 0.183 | V | ⬟ |
| 1.4 | $6 \cdot 10^6$ | 0.15 | 0.94 | 0.0008 | 0.653 | 0.339 | V | ● |
| 2.1 | $1 \cdot 10^7$ | 0.25 | 0.62 | 0.943 | 0.001 | 0.056 | III | ▼ |
| 2.2 | $1 \cdot 10^7$ | 0.25 | 0.67 | 0.912 | 0.0009 | 0.087 | III | ♦ |
| 2.3 | $1 \cdot 10^7$ | 0.25 | 0.72 | 0.0021 | 0.843 | 0.155 | **V** | ■ |
| 2.4 | $1 \cdot 10^7$ | 0.25 | 0.82 | 0.0055 | 0.746 | 0.249 | V | ⬟ |
| 2.5 | $1 \cdot 10^7$ | 0.25 | 0.94 | 0.461 | 0.188 | 0.351 | IV | ● |
| 2.6 | $1 \cdot 10^7$ | 0.25 | ∞ | 0.093 | 0.065 | 0.842 | IV | ✕ |
| 3.1 | $1.5 \cdot 10^7$ | 0.25 | 0.62 | 0.02 | 0.421 | 0.559 | IV | ▼ |
| 3.2 | $1.5 \cdot 10^7$ | 0.25 | 0.67 | 0.033 | 0.435 | 0.533 | V | ♦ |
| 3.3 | $1.5 \cdot 10^7$ | 0.25 | 0.72 | 0.011 | 0.679 | 0.31 | V | ■ |
| 3.4 | $1.5 \cdot 10^7$ | 0.25 | 0.82 | 0.039 | 0.726 | 0.235 | V | ⬟ |
| 3.5 | $1.5 \cdot 10^7$ | 0.25 | 0.94 | 0.558 | 0.322 | 0.121 | **IV** | ● |
| 4.1 | $9 \cdot 10^6$ | 0.15 | 0.62 | 0.014 | 0.625 | 0.361 | V | ▼ |
| 4.2 | $9 \cdot 10^6$ | 0.2 | 0.62 | 0.956 | 0.0012 | 0.042 | III | ▼ |
| 4.3 | $9 \cdot 10^6$ | 0.25 | 0.62 | 0.962 | 0.0005 | 0.037 | **III** | ▼ |
| 5.1 | $1.2 \cdot 10^7$ | 0.25 | 0.72 | 0.027 | 0.716 | 0.257 | V/**VI** | ■ |
| 5.2 | $2 \cdot 10^7$ | 0.25 | 0.72 | 0.024 | 0.437 | 0.539 | V | ■ |
| 5.3 | $2.5 \cdot 10^7$ | 0.25 | 0.72 | 0.018 | 0.472 | 0.510 | V | ■ |
| 6.1 | $9 \cdot 10^6$ | 0.25 | 0.82 | 0.621 | 0.002 | 0.377 | **II** | ⬟ |
| 6.2 | $1.2 \cdot 10^7$ | 0.25 | 0.82 | 0.004 | 0.672 | 0.324 | V | ⬟ |
| 6.3 | $2 \cdot 10^7$ | 0.25 | 0.82 | 0.014 | 0.563 | 0.423 | V | ⬟ |
| 6.4 | $3 \cdot 10^7$ | 0.25 | 0.82 | 0.042 | 0.267 | 0.691 | V | ⬟ |
| 7.1[5] | $1 \cdot 10^7$ | 0.25 | 0.62 | 0.685 | 0.006 | 0.309 | II | ▼ |
| 7.2[5] | $1 \cdot 10^7$ | 0.25 | 0.72 | 0.001 | 0.667 | 0.332 | V | ■ |
| 7.3[5] | $1 \cdot 10^7$ | 0.25 | 0.94 | 0.177 | 0.207 | 0.614 | IV | ● |
| 8.1† | $5 \cdot 10^6$ | 0.15 | 0.62 | 0.287 | 0.299 | 0.414 | **VII** | ▽ |
| 8.2† | $9 \cdot 10^6$ | 0.25 | 0.62 | 0.761 | 0.005 | 0.234 | III | ▼ |
| 8.3† | $9 \cdot 10^6$ | 0.25 | 0.94 | 0.122 | 0.072 | 0.806 | IV | ● |
| 8.4† | $1 \cdot 10^7$ | 0.25 | 0.94 | 0.118 | 0.091 | 0.791 | IV | ● |
| 9 | $7 \cdot 10^6$ | 0.15 | 0.62 | 0.024 | 0.558 | 0.439 | IV | ▼ |
| 10 | $7 \cdot 10^6$ | 0.2 | 0.67 | 0.945 | 0.001 | 0.054 | III | ♦ |
| 11 | $1 \cdot 10^7$ | 0.2 | 0.67 | 0.473 | 0.225 | 0.302 | III/IV | ♦ |
| 12 | $8 \cdot 10^6$ | 0.15 | 0.72 | 0.898 | 0.008 | 0.095 | III | ■ |
| 13♯ | $1.1 \cdot 10^7$ | 0.1 | 0.94 | 0.874 | 0.004 | 0.128 | **I** | ● |

It is interesting to note that we have not encountered any bistability in our runs. Bistability is the coexistence of two (or more) distinct stable solutions, where the emerging solution in any particular run being determined by the initial conditions. Bistability was reported for similar models by Duarte et al. (2013), but typically only for Jupiter-like models with rather high $r_d$. In the scenario of Duarte et al. (2013), either a strong, dipolar field alongside weak zonal flows, or a multipolar field shredded by the strong zonal flows, are the two distinct characteristic solution types. For our models, especially those with small $r_d$, zonal flows are always strongly generated in the hydrodynamic shell, where they cannot be attenuated by the magnetic field. Furthermore our models operate at a slightly different parameter regime. Bistability cannot be ruled out for our models, but it seems uncommon in the range of models we examined.

## 4. Results

For an overview, we consider models where only the radial position of the conductivity drop-off $r_d$ is changed, but all other parameters are kept identical. To compare the emerging solutions to end-member scenarios, a pure hydrodynamic simulation with no magnetic fields and a fully conducting model where the electrical conductivity is constant along radius are added. The details of these runs can be found in group 2 of Table 1. After time integrating through a transient we average the solutions over a fraction of the magnetic diffusion time, and some results are shown in Fig. 3. Zonal flows are consistently found to be prograde at the equator regardless of the magnetic effects (first column in Fig. 3 and the surface snapshots), but obey rather different amplitudes and patterns, discussed below in section 3.1. In the second column of Fig. 3 we see that the meridional circulation is small in these rapidly rotating low Rossby number models, but there is some meridional flow coupling the magnetic conducting interior to the non-magnetic exterior. The third column gives the helicity, discussed further in section 4.2, an important quantity for magnetic field generation. The magnetic fields are shown in the fourth and fifth columns of Fig. 3), along with a snapshot of the surface radial field. It is immediately apparent that the morphology of the magnetic field is remarkably different for the different values of $r_d$, even though all other parameters are identical. In particular, between the Jupiter-like case (b) and the Saturn-like case (d) there is a region of $r_d$ where the field is quadrupolar rather than dipolar, i.e. the radial magnetic field is symmetric rather than antisymmetric about the equator. We discuss these differences in sections 4.3–4.5 below.

### 4.1. Zonal flows and the conservation of angular momentum

The resulting surface zonal flows for the set of models where only the conductivity drop-off $r_d$ is changed (group 2 in Table 1) are plotted in Fig. 4. The plot shows the time-averaged, axisymmetric azimuthal flow ($\bar{u}_\phi$) at the surface as a function of latitude in terms of the Rossby number $Ro_s = \bar{u}_\phi E(1-\beta)/Pm$. The horizontal dashed lines denote the virtual magnetic tangent cylinder and are coloured according to the $r_d$-value used. It can be seen that the model with smallest $r_d$ has a prograde equatorial peak jet amplitude of more than double the Jupiter-like models (Fig. 4 orange vs green profile). The broader and more energetic jet is in line with the observed profiles for Jupiter and Saturn (Porco et al., 2003; Sanchez-Lavega et al., 2000; Aurnou et al., 2007) with peak equatorial amplitudes of 150 and 450 m/s, respectively. For completeness the fully conducting model ($r_d = \infty$) and a hydrodynamic model without magnetic field are also included (purple and grey profile). In the fully conducting model, the equatorial peak jet is as small as $Ro_e = 0.011$ hence only a quarter of the Jupiter-like model with $r_d = 0.94$. This indicates that even a rather thin non-magnetic shell creates a firm equatorial jet. Further as the difference between a deep drop-off model ($r_d = 0.62$) and the hydrodynamic run ($Ro_e = 0.08$ vs. 0.11) is much weaker, it is clear that the zonal flow system originates mainly from the outer regions.

The main force balances are fundamentally different between the hydrodynamic outer and magnetic inner shell. In the former, zonal flows are exclusively maintained by Reynolds stress created by a consistent strong tilt in the convective columns and ultimately saturated by the fluid viscosity. This leads to strong geostrophic zonal flows. However, in the metallic region the Lorentz force counteracts the rotational forces, and the convective columns are stopped from further tilting by magnetic tension. Hence the differential rotation emerging inside the magnetic tangent cylinder is considerably weaker. This explains why the models with only a thin non-magnetic shell (or no non-magnetic shell) have a weaker zonal flow, and why the models with a smaller $r_d$ have a broader prograde jet. However, the central prograde jet on Saturn (Jupiter) extends to $\pm 35°$ ($\pm 20°$), whereas in our results the equatorial jets are broader. Also, despite the reasonably low Ekman number ($E = 5 \times 10^{-5}$) there is not much evidence for any significant westward jets, which are known to exist in higher latitude regions on Jupiter and Saturn. It is possible that at much smaller $E$ westward jets will develop, but it does appear that while westward jets can be readily obtained in purely hydrodynamic models (e.g. Jones and Kuzanyan, 2009), it is much harder to get westward jets in models with a magnetic field.

Fig. 4 clearly indicates the importance of applying a radially varying electrical conductivity in attempts to model gas giant atmospheric circulations. As already suggested in Jones (2014), our results indicate that there are deep jets outside the magnetic tangent cylinder, giving rise to the large surface flows near the equatorial region, but that the jets at higher latitudes, which lie inside the magnetic tangent cylinder, are due to localised surface effects in the uppermost parts of giant planet atmospheres.

To more clearly identify the main force balance maintaining the differential rotation system two equations are helpful. The first is the thermal wind equation, which is the $\phi$-component of the vorticity equation, so we take the time-averaged $\phi$-component of the curl of equation (6),

$$\omega_\phi (\nabla \cdot \boldsymbol{u}) + s\,\boldsymbol{u} \cdot \nabla\left(\frac{\omega_\phi}{s}\right) - s\boldsymbol{\omega} \cdot \nabla\left(\frac{u_\phi}{s}\right) = \frac{2Pm}{E}\frac{\partial u_\phi}{\partial z}$$
$$-\frac{RaPm^2}{Pr}\frac{1}{r}\frac{\partial}{\partial \theta}\left(S\frac{d\bar{T}}{dr}\right) + \frac{Pm}{E}\left(\nabla \times \left(\frac{1}{\bar{\rho}}(\nabla \times \boldsymbol{B}) \times \boldsymbol{B}\right)\right)_\phi$$
$$+Pm(\nabla \times \boldsymbol{F}_\nu)_\phi, \tag{15}$$

where $s = r\sin\theta/r_o$ is the distance from the rotation axis and $\boldsymbol{\omega} = \nabla \times \boldsymbol{u}$ is vorticity. In low Rossby number giant planets, the nonlinear inertia terms on the left-hand-side are much smaller than the first term on the right, since planetary vorticity dominates local vorticity, and the final viscous term is also small. The thermal wind equation is most useful outside the magnetic region, $r > r_d$, where the magnetic terms can be neglected. We are then left with the usual thermal wind balance between latitudinal entropy gradients and the gradient of $u_\phi$ parallel to the rotation axis (z-axis). However, entropy gradients are small in the convective regions of giant planets because efficient convection ensures that the whole atmosphere is close to adiabatic. In consequence of $\partial u_\phi/\partial z = 0$, $u_\phi$ has to be nearly independent of $z$, i.e. geostrophic. Note that this argument breaks down in the Sun, because the Rossby number in the convection zone is not very small, so the terms on the left are non-negligible, but the convection in giant planets is driven only by very slow cooling, and so is much weaker. It is also possible that close to the surface the density is low enough for significant entropy (hence temperature) fluctuations to occur, so the flow there could be ageostrophic, but our model has to be cut off before such very low densities are reached. It is now clear why there is so little zonal flow inside the tangent cylinder even in the non-metallic regions: the Maxwell stresses wipe out any large zonal flows in the metallic region, and this sets a $u_\phi = 0$ bottom boundary condition for the thermal wind equation (15). Since there are no significant terms forcing a thermal wind gradient, it remains close to zero everywhere inside the magnetic tangent cylinder. Outside the magnetic tangent cylinder, there is no equivalent bottom boundary condition, and large geostrophic zonal flows can build up as foreseen by Busse (1983).

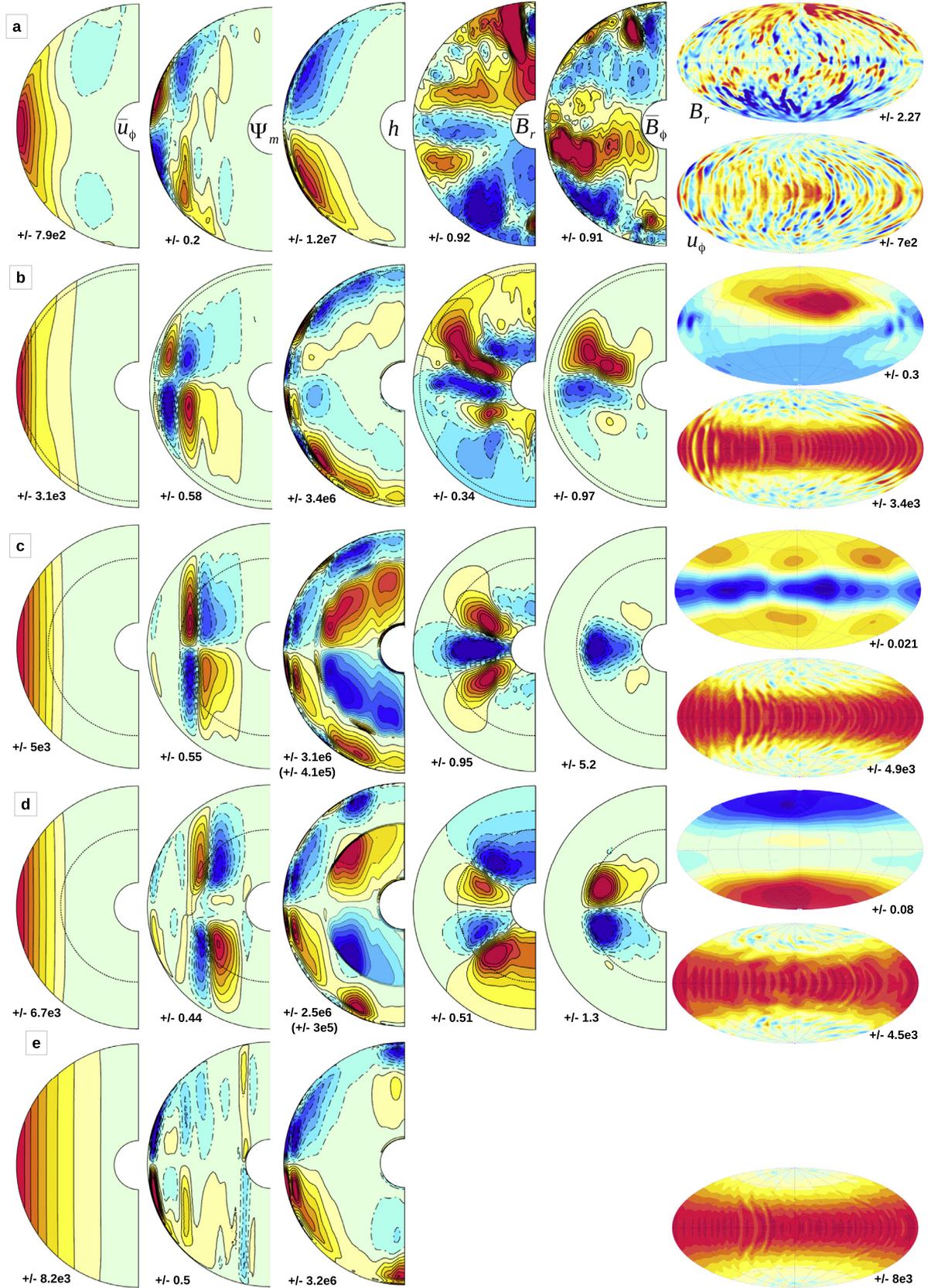

**Fig. 3.** From left to right we show five time-averaged meridional sections. These are: the azimuthal flow ($\bar{u}_\phi$), the meridional stream function $\Psi_m$, the kinetic helicity $h$, the radial and azimuthal fields $\bar{B}_r$ and $\bar{B}_\phi$. Also shown are typical snapshots of the spherical projection of the radial field and azimuthal flow at the surface. From top to bottom the five cases are: (a), the fully conducting model, run 2.6 in Table 1; (b) the Jupiter-like model, run 2.5, with $r_d = 0.94$; (c) an intermediate $r_d = 0.72$ model, run 2.3; (d) the deep drop-off case $r_d = 0.62$, run 2.1; (e) the hydrodynamic model. In (b–d), the dashed curve gives the approximate radial level where the electrical conductivity drops off. The maximal contour levels are listed. For the helicity in figures (c) and (d), the helicity in the interior is amplified for clarity, and the maximal contour levels in the interior region are given in brackets. Flows are in terms of magnetic Reynolds numbers $Rm$, the field in measures of the Elsasser number $\Lambda$. Parameters: $Ra = 1 \cdot 10^7$, $E = 5 \cdot 10^{-5}$, $Pr = 0.25$, $Pm = 3$.

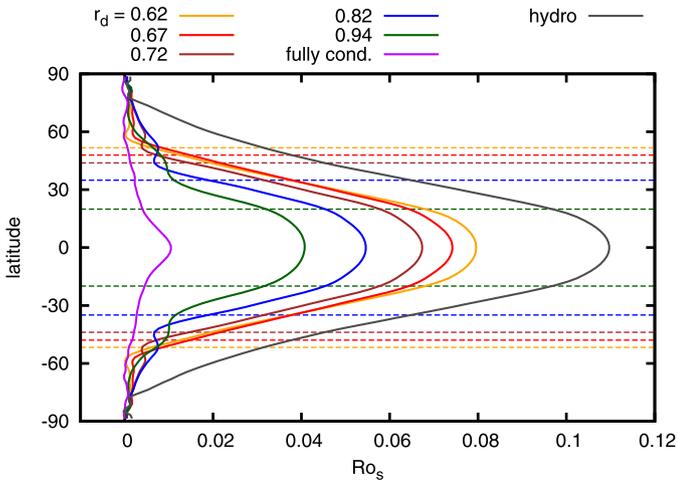

**Fig. 4.** Axisymmetric surface zonal flow along latitude for models with variable conductivity drop-off $r_d$, a fully conductive model (magenta) and pure hydrodynamic simulation (dark-grey). The horizontal dashed lines denote the latitude. of the virtual magnetic tangent cylinder attached at $r_d$.(For interpretation of the references to colour in this figure legend, the reader is referred to the web version of this article.)

The second useful equation is that governing zonal angular momentum per unit mass $L$

$$L = \bar{u}_\phi s + \frac{Pm}{E} s^2. \quad (16)$$

The first term gives the differential rotation and the second concerns the planetary solid body rotation. The conservation equation can be found by multiplying the azimuthal component of the Navier–Stokes equation (6) by $s$ and averaging over azimuth and time (e.g, Browning, 2008; Schneider and Liu, 2009; Liu and Schneider, 2010; Gastine et al., 2013):

$$\overline{\rho}\,\bar{\mathbf{u}}_m \cdot \nabla L = \nabla \cdot \left[ Pm\overline{\rho} s^2 \nabla \left( \frac{\bar{u}_\phi}{s} \right) - \overline{\rho} s \overline{\mathbf{u}'u'_\phi} \right]$$
$$+ \nabla \cdot \left[ \frac{Pm}{E\overline{\rho}} s \left( \overline{B'B'_\phi} + \overline{B}\,\overline{B}_\phi \right) \right]. \quad (17)$$

where the remaining forces due to viscosity, Reynolds stress and Maxwell stress appear in that order on the right hand side of the equation.

This diagnostic equation links the axisymmetric flows (meridional circulation $\bar{\mathbf{u}}_m$ and zonal flows $\bar{u}_\phi$) with the non-axisymmetric convection $\mathbf{u}'$ and the magnetic field contributions $\overline{B}\,\overline{B}_\phi$ in terms of forces due to viscous, Reynolds and Maxwell stresses, respectively. The left hand side of Eq. 17 contains only mean flows and hence is easily calculated, whereas on the right hand side, the terms due to Reynolds and Maxwell stress require correlating a set of snapshots over time and azimuth. We verified that for each case the left and right hand sides are equal.

Fig. 5 plots the five terms for the same set of models as in Fig. 3 (group 2 in Table 1). In the hydrodynamic case (Fig. 5(e)), the viscosity balances the Reynolds stresses to a large extent, hence the advection of zonal angular momentum $L$ by meridional circulation is marginal. This is in line with the classic picture of the zonal flow created by a consistent tilt in the convective columns which further enhance the differential rotation (Busse, 2002). This run-away process is ultimately stopped by the viscosity acting on the shear of the zonal flows. In contrast, in a fully conducting model (Fig. 5(a)) the Reynolds stresses are stopped from further growth by the magnetic forces and the viscosity is negligible (Aubert, 2005). As a consequence, the zonal flows are much weaker if magnetic fields are invoked as already shown in Fig. 4. Note for both, the fully conducting and the hydrodynamic model (Fig. 5(a) and (e)), the balance between the Reynolds and either Maxwell or viscous stresses is fairly well fulfilled, such that the advection of angular momentum by meridional circulation is marginal.

However, when the electrical conductivity is a function of radius (Fig. 5(b)–(d)) and hence a hydrodynamic shell encloses a magnetic deeper interior, there we find contributions from all three stress sources. The Maxwell stresses are confined within the metallic region, hence cannot balance the Reynolds stresses which are quite strong in the non-metallic region. To achieve a z-invariant differential rotation a mean advection of angular momentum is required to maintain the balance. Note the strong Reynolds stresses in the 3rd column of Fig. 5(d) that occur near the outer boundary just inside the mTC. The thermal wind constraint means that these Reynolds stresses cannot drive a zonal flow, so they have to be almost exactly balanced by a mean advection of angular momentum, see the 1st column of Fig. 5(d).

In the metallic region, the dominant balance seems to be between the Maxwell stresses (columns 4 and 5) and the advection of angular momentum (column 1). The Reynolds stresses are not completely negligible, but they have a minor role there. Near the equatorial plane, zonal angular momentum is transported inwards from the hydro region (blue areas in the figure in the first column) and outwards in the deep metallic region (red). These converging flows are directed polewards to conserve the mass flux and lead to the characteristic meridional circulation seen in Fig. 3. Although the zonal flow itself is not large near this convergent region, its gradient gives rises to a noticeable viscous contribution in some of the 2nd column panels of Fig. 4, though this will likely disappear at lower $E$. It is clear that the zonal force balance is rather complex for systems coupling hydrodynamic and magnetic regions and can include all three forces plus meridional circulation.

### 4.2. Kinetic helicity

The kinetic helicity $h$ is thought to play an essential part in the generation of magnetic fields in rapidly rotating convection (Moffatt, 1978; Olson et al., 1999; Sreenivasan and Jones, 2011). Fig. 3 third column plots the azimuthal average of the kinetic helicity gained from non-axisymmetric flows:

$$h = \overline{\mathbf{u}' \cdot (\nabla \times \mathbf{u}')}. \quad (18)$$

We time-average the helicity over a series of snapshots and find the classical picture of an antisymmetric distribution with a negative (positive) mean in the northern (southern) hemisphere for the fully conducting (top) and hydrodynamic model (bottom). This is the usual parity found in Boussinesq models. We note that, in agreement with previous findings (Sreenivasan and Jones, 2011), the kinetic helicity in the fully conducting magnetic case is three times higher than in the hydrodynamic one. This is in line with the classic linear picture of z-independent convective columns ex-tending along the height of the entire shell. The negative helicity stems from a consistent alignment of flow and flow vorticity, hence the vertical flow is equatorward in a cyclonic column and poleward in an anticyclone. The convective structures in our simulations are not strictly columnar due to their nonlinearity, but the dominant Coriolis force still enforces much larger vertical than horizontal length scales.

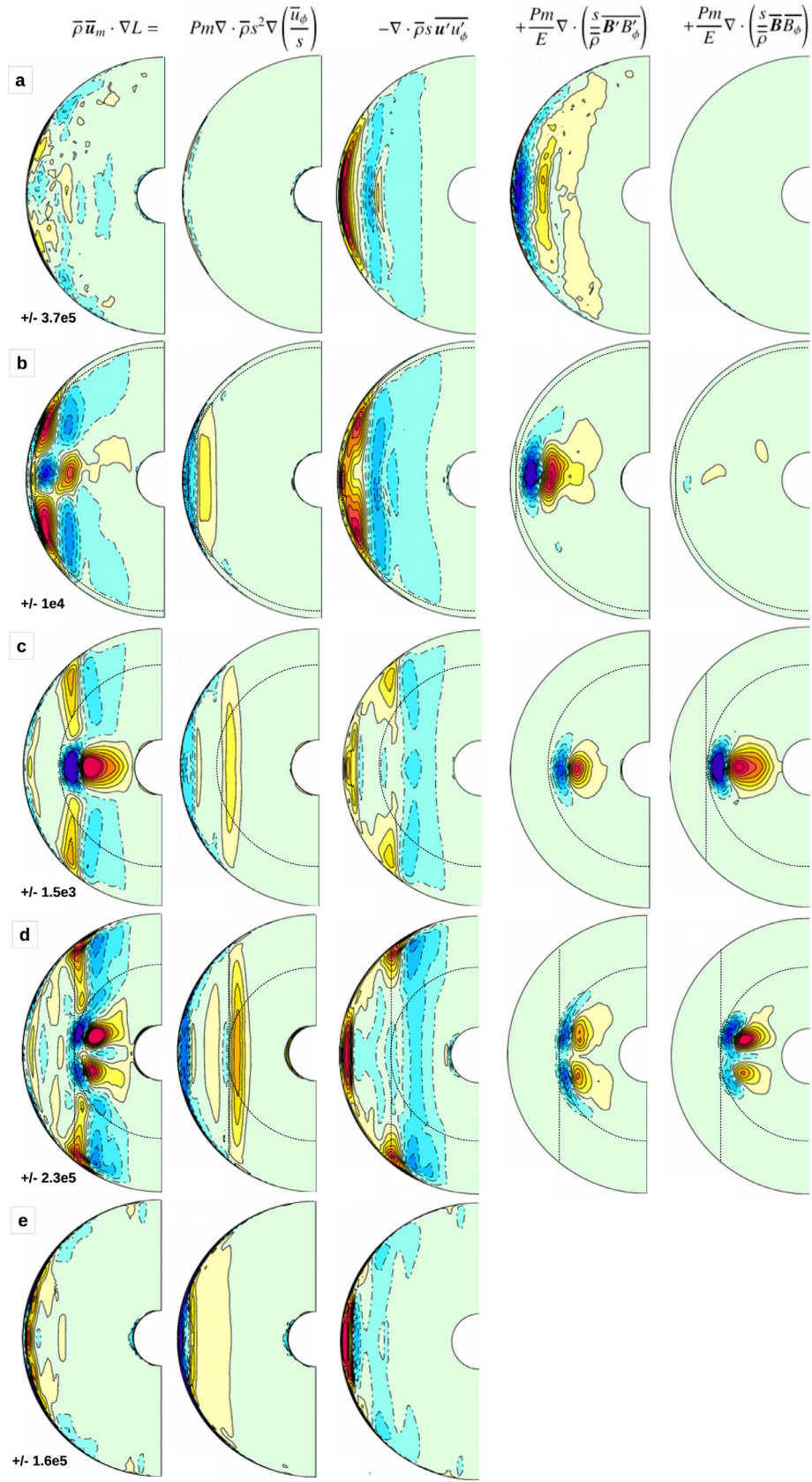

**Fig. 5.** Time-averaged balance of angular momentum. From left to right: angular momentum flux is given by forces due to viscous, Reynolds, non-axisymmetric and axisymmetric Maxwell stresses. From top to bottom: fully conducting model (a), the Jupiter-like model with $r_d = 0.94$ (b), intermediate $r_d = 0.72$ (c), deepest drop-off $r_d = 0.62$(d) and hydrodynamic model (e). The dashed semi-circles denote the conductivity drop-off, the vertical dashed lines the corresponding magnetic tangent cylinders. The runs presented are the same as in Fig. 3.

For the hydrodynamic model (Fig. 3, bottom) it can be seen that inside the tangent cylinder, both helicity parities are visible in each individual hemisphere. Near the rotation axis, radial upwards motion is accompanied by a divergence of horizontal flow which is turned into a cyclonic structure by the Coriolis force and vice versa. Note that the helicity sign change is due to a flip of the vorticity, whereas for the columnar mode the kinetic helicity flips at the equator due to a change in the vertical flows (Guervilly et al., 2014; Duarte et al., 2016).

The remarkable feature of the third column of Fig. 3 is the reversed helicity in the metallic region, positive in the northern hemisphere and negative in the southern hemisphere. Anelasticity allows for vorticity generation by the background density gradient. This will create negative vorticity in rising, expanding fluid parcels and hence negative helicity in the northern hemisphere. This aligns with the helicity distribution predicted for (also incompressible) columnar convection, and so would simply enhance the usual Boussinesq helicity distribution. It has been suggested that internal heating and strong inertia effects can lead to inverse helicity, i.e. positive in the north and negative in the southern hemisphere (Duarte et al., 2016). However, our hydrodynamic model shows the usual behaviour, so we attribute the inverse helicity patches found in models with variable conductivity (Fig. 3 middle three panels) to an effect of the magnetic field.

As we saw in the zonal flow section, magnetic fields are dynamically important, so we expect them to affect the helicity. However, given that the fully magnetic model shows the original helicity distribution found for the simpler hydro-model, it is clear that helicity inversion is not simply due to the magnetic field itself. Duarte et al. (2016) argue that the regions close to the outer boundary and the deeper interior can be decoupled by a large density gradient. Apparently in our models it is the electrical conductivity and hence the magnetic field which separates the two distinct regions. Closer to the outer boundary, the helicity distribution is controlled by the boundaries, hence axial equatorward (upward) flow is associated with cyclones (anticyclones) leading to the classical picture of negative kinetic helicity in the northern hemisphere. However in the magnetic regions, the Lorentz force is of first order and relaxes the rotational constraints. Hence the axial vorticity is no longer controlled by (rather distant) boundaries but by the diverging/converging flows near the equatorial plane driven by the magnetic field. This leads to positive kinetic helicity as axial equatorward flows are now apparently associated with anticyclones and vice versa. As the inversion consistently emerges for models with variable conductivity it can be concluded that it requires different force regimes to decouple the boundary control of the helicity parity. This can be achieved by e.g the combination of hydrodynamic and magnetic regions with strong density gradient in the outermost layers (Duarte et al., 2016).

Sreenivasan and Jones (2011) found that quadrupolar fields reduced the helicity while dipolar fields enhanced it, and attributed the overall preference of dipolar fields in Boussinesq models to this effect. In none of their models was the helicity ever reversed. It appears that in these variable conductivity models, the combination of magnetic field and induced mean flows is having such a strong effect that both the quadrupolar Fig. 3(c) and the dipolar Figs. 3(b) and (d) undergo helicity reversal. It is therefore perhaps not surprising that these anelastic variable conductivity models show no strong preference between dipolar and quadrupolar fields.

### 4.3. Dynamo solutions

As we observe many different types of magnetic field in our solutions, we firstly aim to distinguish different temporal behaviour by plotting butterfly diagrams (Fig. 6). The parameters of the runs are listed in Table 1. These plots show the axisymmetric, radial field at the surface of the spherical shell as a function of colatitude and time. For the Jupiter-like models with a thin non-conducting region, a dominant stable dipole is a possible solution, in agreement with the observations of Jupiter's magnetic field (panel a). Our example in panel (a), model 13, is close to the Jupiter model of Jones (2014). In our group 2 models with $r_d = 0.94$ we also found a dominant dipolar field case (Table 1, model 2.5), but this model harbours a more erratic temporal evolution with some infrequent polarity reversals. When the non-conducting outer shell is slightly increased, so $r_d = 0.82$, model 6.1 in Fig. 6(b), the octupole participates strongly, leading to positive and negative radial magnetic flux in both hemispheres. However, when the conducting region is shrunk further, $r_d = 0.62$, and all other parameters are kept the same, the dipole is again dominant, but it starts to oscillate, see Fig. 6(c) model 4.3. The field has the form of a dynamo wave propagating towards the equator, as in the Sun. Although the differential rotation is primarily in the non-conducting region, some differential rotation does penetrate into the outer parts of the dynamo region (see the leftmost column of Fig. 3(c) and (d)), and we believe this provides an omega-effect giving rise to dynamo waves. Note that Fig. 6(c) has a slightly smaller Rayleigh number than Figs. 6(a) and (b).

In Fig. 6(d) (model 3.5), we show what happens if the Rayleigh number is significantly increased. The dynamo becomes small scale and irregular. This example is for $r_d = 0.94$, but the same happens at all $r_d$ if $Ra$ is increased sufficiently. This is expected from experience with Boussinesq models, as the more vigorous convection is energetic enough to overcome the dominant geostrophy (local Rossby number no longer sufficiently small) and hence irregular, small-scale fields are created by more isotropic turbulence.

However, a more moderate increase of the vigour of convection ($Ra$) can turn the dynamo wave behaviour of Fig. 6(c) into a steady quadrupole solution, as shown in Fig. 6(e), model 2.3. Although this model has a magnetic field that is almost completely symmetric about the equator, dipole and quadrupole modes are not completely exclusive. Fig. 6(f), model 5.1, is at a slightly higher Rayleigh number, and although it is basically a quadrupolar dynamo, it has a small oscillatory dipolar component as well. Such a combination of an oscillating dipole and a steady quadrupole seems to be the most common solution for models with intermediate $r_d$ studied in this paper.

As a peculiar solution, we also identify a hemispherical wave dynamo (Fig. 6(g)), model 8.1. The Rayleigh number is moderate, $r_d = 0.62$ and $Pr$ is smaller, at 0.15 rather than 0.25. This solution is remarkable as dipolar and quadrupolar contributions must be equally strong and are required to oscillate in a well-defined phase relation in order to keep the field hemispherical at all times. For all the other solutions it seemed that dipolar and quadrupolar solutions are created independently of each other and are simply superimposed. This quite obviously cannot be true for the hemispherical wave. Note that the magnetic field there is weaker by a factor 10 compared to the other solutions.

In Fig. 7 we show some meridional sections of the hemispherical dynamo wave, which shed light on the nature of this solution. First we note from row (a) the latitudinal gradients of the entropy, are quite equatorially symmetric, so the small thermal wind is also symmetric about the equator, and consistent with this the zonal flow is also symmetric. Also the kinetic helicity distribution is almost perfectly antisymmetric to the equator. Thermal wind, zonal flow and the kinetic helicity as a proxy for the induction of poloidal field are therefore not the origin of the hemispherical nature of the dynamo. It is the meridional circulation which shows a strong equatorial asymmetry. The reason for this can be seen in the angular momentum flux balance shown in row (b). The viscous forces and the Reynolds stresses are quite equatorially symmetric, but the Maxwell stresses are not, as we would expect since the magnetic field resides primarily in one hemisphere. Inside the metallic region, the main balance is between the Maxwell stresses and the advection of angular momentum by the meridional circulation, so this forces the meridional circulation to be equatorially asymmetric. This suggests that a hemispherical dynamo comes about when an asymmetric meridional circulation pushes the azimuthal magnetic field slightly off centre, and this offset magnetic field then generates the right form of meridional circulation that advects the magnetic field still further off centre. In the dynamos which remain in either dipolar or quadrupolar parity, presumably a small offset generates a meridional circulation which corrects, rather than exacerbates, the off centre components of the magnetic field.

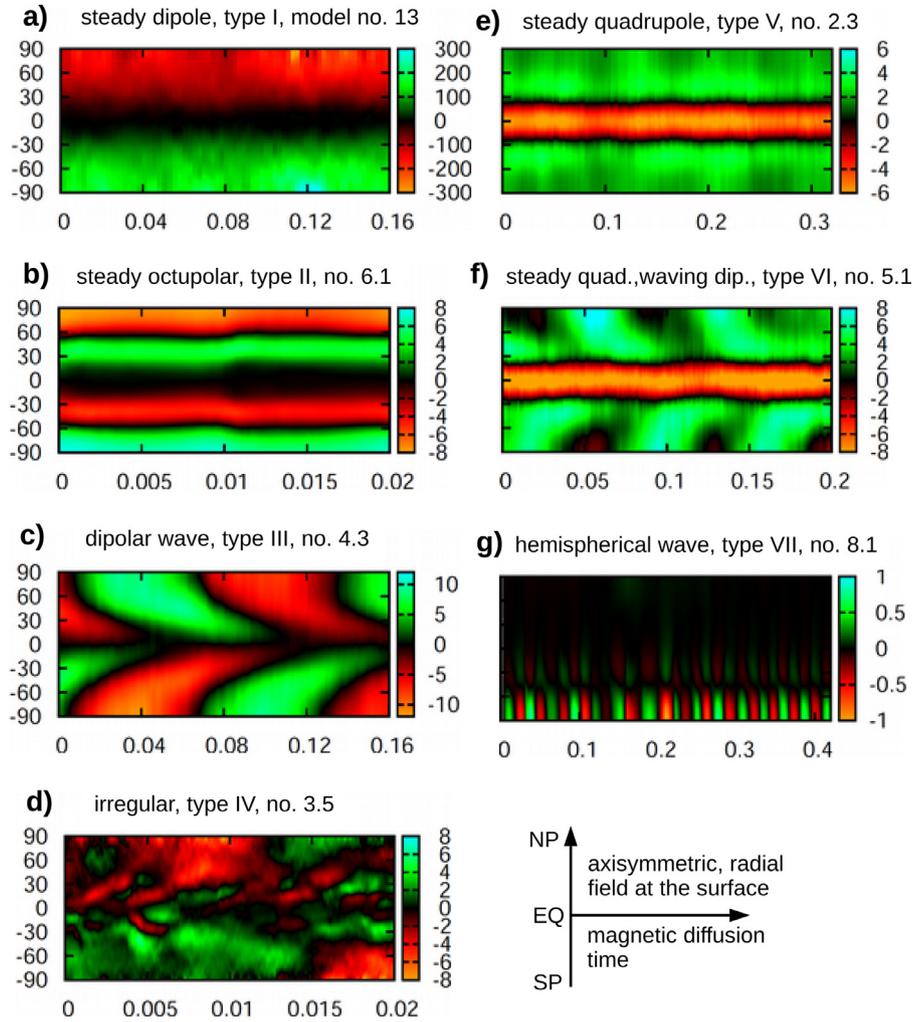

**Fig. 6.** Axisymmetric radial field at the outer boundary as function of latitude and time (butterfly diagram) to characterise the distinct solutions. Details of the input parameters are given in Table 1.(For interpretation of the references to colour in this figure legend, the reader is referred to the web version of this article.)

To give further information about the nature of the whole range of our solutions, we show the phase relation between dipole and quadrupole components in Fig. 8 by plotting a phase diagram of the surface values of $g_{20}$ against $g_{10}$ as time evolves ($g_{10}$ and $g_{20}$ are defined in equation (20) below). The numbers (I-VII) indicate for each trajectory the various solution types highlighted in Table 1 and correspond to the cases shown in Fig. 6. This plot is sensitive to sign changes, but it only covers two single axisymmetric spherical harmonics out of the possibly complex magnetic field. The diagonal dashed lines are indicating $g_{10} = \pm g_{20}$, and the hemispherical dynamo VII operates along this line, as equal contribution of $g_{20}$ and $g_{10}$ are needed to minimise the field in one hemisphere and maximise it in the other at all times. The stable dipolar dynamo I (in blue), which is Jupiter-like, hovers around $g_{10} = 0.01$ and $g_{20} = 0$, as there is some variation of the field strength with time even though the sign of $g_{10}$ doesn't change. The solar-like dipolar waves, solution III (dark-red) show only a small quadrupolar contribution and oscillate along the $g_{10}$-line. The path for solution IV (red), the small scale irregular dynamo, wanders about erratically as expected. The orange trajectory, solution V, corresponding to the quadrupole, has only a small dipolar term and stays around $g_{20} = 0.006$. The quadrupolar dynamo with an oscillatory dipole component, solution VI in green, wanders more in the dipolar direction, but interestingly never crosses the $g_{10} = \pm g_{20}$ diagonals, suggesting that in these models dipolar and quadrupolar modes do depend on each other. In other words, increasing the convective vigour leads to an increased strength of the dipole, but it never exceeds the quadrupole. The peak magnetic intensity then oscillates between north and south.

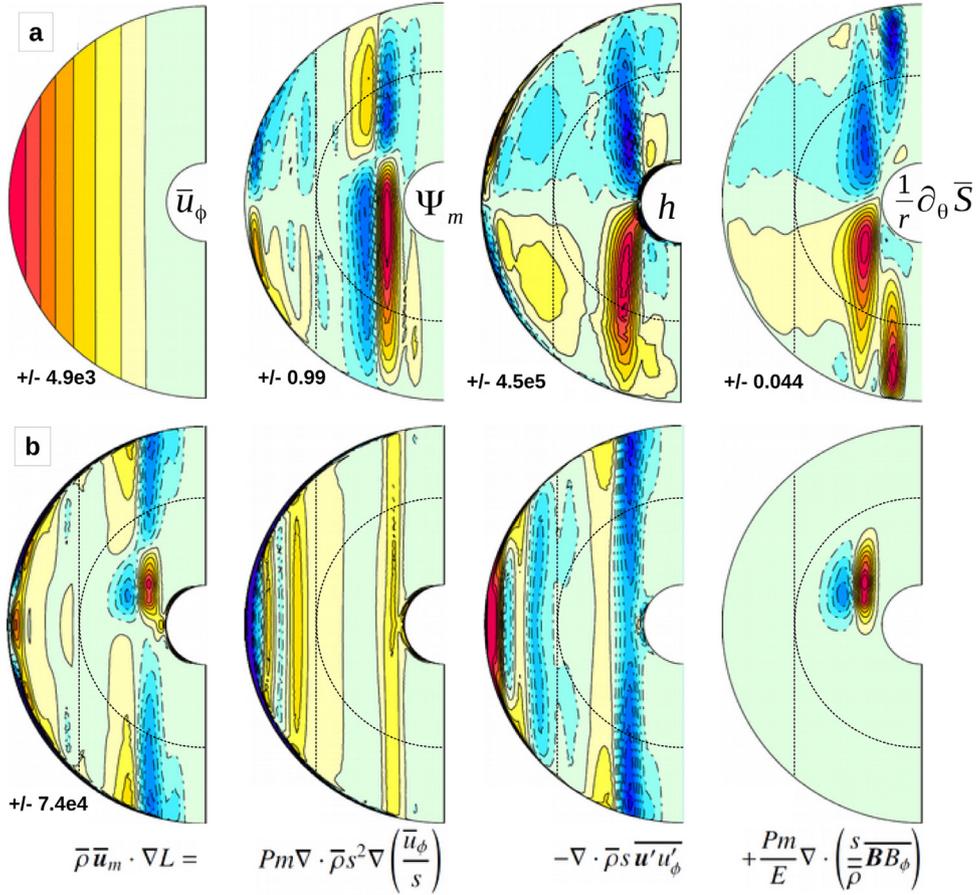

**Fig. 7.** Time-averaged meridional sections of the hemispherical wave dynamo, model 8.1. (a) from left to right: the azimuthal flow ($\bar{u}_\phi$), the meridional stream function $\Psi_m$, kinetic helicity $h$ and the latitudinal entropy gradient $\frac{1}{r}\partial_\theta \bar{S}$. (b) the angular momentum flux balance, as in the first 4 panels of Fig. 5.

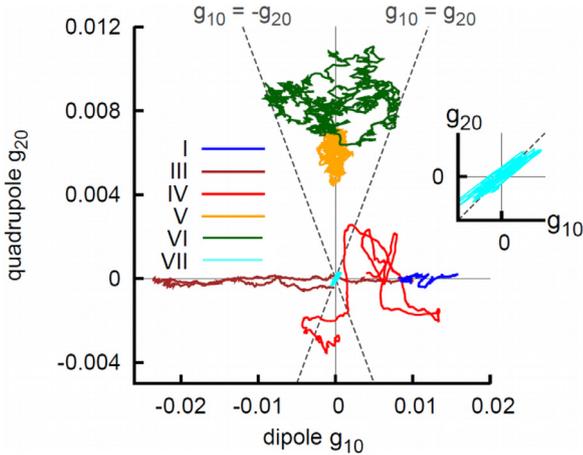

**Fig. 8.** Phase diagram $g_{20}(g_{10})$ to identify the various regimes. $g_{10}$ and $g_{20}$ are the sign and amplitude of the axisymmetric dipole and quadrupole field component at the surface. The small inset is an upscaled plot of the hemispherical wave with weaker magnetic amplitude indicating the strict phase relation between dipole and quadrupole. The roman numbers refer to the types in Table 1 and Fig. 6.

### 4.4. Magnetic trigram

The 'butterfly diagrams' are helpful for identifying the time-dependence of the axisymmetric radial field at the surface, which is important for giving an overall view of the magnetic field, but they do not tell us about the non-axisymmetric components of the field. It is therefore helpful to measure the relative axisymmetry of our solutions, to properly compare the solutions to satellite measurements of real planetary magnetic fields. This distinguishes between highly axisymmetric planets like Saturn and those with substantial non-axisymmetric components like Uranus and Neptune. The observed fields are typically characterised by a set of 'Gauss coefficients' ($g_{\ell m}$, $h_{\ell m}$), each giving the amplitude of a specific spherical harmonic contribution to the internal field. Assuming that the dynamo field field created inside the planet can be upward-continued through a source-free region, the resulting magnetic field can be expressed as the gradient of a scalar potential $V$:

$$\mathbf{B} = -\nabla V , \qquad (19)$$

where $V$ is constructed with the Gauss coefficients

$$V = r_p \sum_{\ell=1}^{\infty} \left(\frac{r_p}{r}\right)^{\ell+1} \sum_{m=0}^{\ell} P_\ell^m(\cos\vartheta)\left\{g_{\ell m}\cos(m\phi) + h_{\ell m}\sin(m\phi)\right\} . \qquad (20)$$

Here $r_p$ is the surface radius of the planet, $P_\ell^m$ are the Legendre polynomials of degree $\ell$ and order $m$ and $g_{\ell m}$, $h_{\ell m}$ the individual Gauss coefficients. An overview of the sets of Gauss coefficients measured for the solar system planets can be found in Connerney (2007). Out of the Gauss coefficients the relative contribution of various symmetries, such as axisymmetry and equatorial symmetry can be extracted even if only a few coefficients are known. Therefore we separate the magnetic energy at the outer boundary contained in axisymmetric and non-axisymmetric modes, where the first group is further subdivided by equatorial

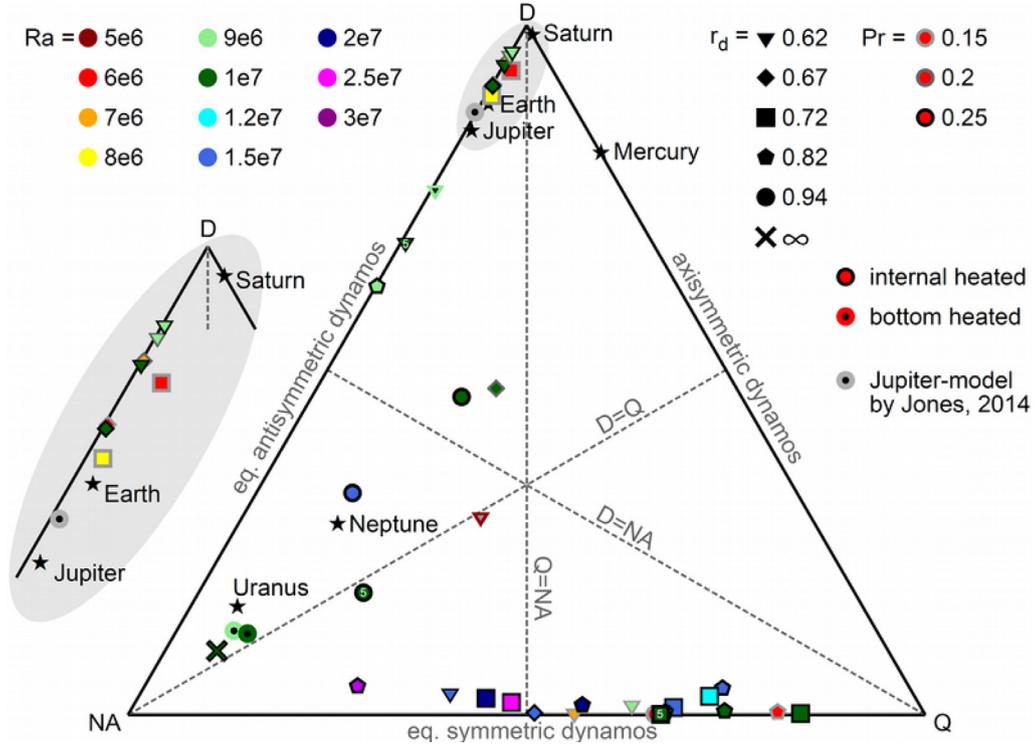

**Fig. 9.** Magnetic trigram for the solar system planets and all numerical models. D/Q refer here to the magnetic energy kept in axisymmetric and equatorially antisymmetric/symmetric modes relative to the total magnetic energy at the planetary surface/the upper boundary of the simulation domain. NA are the remaining non-axisymmetric contributions. Colours refer to the Rayleigh number, symbol shapes to the radial position of the conductivity drop-off and the grey-scaled edge indicates the Prandtl number. Note, small '5' indicates models with higher magnetic Prandtl $Pm = 5$, whereas $Pm = 3$ otherwise. Further the few models with powered by a fixed entropy contrast and no internal heat source are characterised with coloured edge and grey inlay. The uppermost trigram domain close to 'D' is amplified by a factor 3 at the left hand side. The Ekman number is fixed at $E = 5 \cdot 10^{-5}$. The single grey dot represents model 13, the Jupiter-model by Jones (2014) with different parameters ($E = 2.5 \cdot 10^{-5}$, $Ra = 9 \cdot 10^6$ and $Pr = 0.1$).

symmetry,

$$D = \sum_{\ell=1}^{\ell_{max}} (\ell+1) g_{\ell 0}^2 / E_m^{tot}, \quad \text{for } \ell \text{ odd, } m = 0,$$

$$Q = \sum_{\ell=2}^{\ell_{max}} (\ell+1) g_{\ell 0}^2 / E_m^{tot}, \quad \text{for } \ell \text{ even, } m = 0,$$

$$NA = \sum_{\ell=1}^{\ell_{max}} \sum_{m=-\ell}^{\ell} (\ell+1) \left(g_{\ell m}^2 + h_{\ell m}^2\right) / E_m^{tot}, \quad \text{for } m \neq 0,$$

$$E_m^{tot} = \sum_{\ell=1}^{\ell_{max}} \sum_{m=-\ell}^{\ell} (\ell+1) \left(g_{\ell m}^2 + h_{\ell m}^2\right), \quad \text{for all } \ell, m,$$

(21)

where $E_m^{tot}$ is the total (poloidal) magnetic energy at the outer boundary. In short, D contains all relative contributions of equatorially antisymmetric and axisymmetric modes, whereas Q incorporates all equatorially symmetric and axisymmetric modes and NA is the deviation from symmetry along the rotation axis. Hence summing up the three parts yields

$$D + Q + NA = 1. \quad (22)$$

This can be represented uniquely in a triangular plot. Fig. 9 shows the *magnetic trigram* which has been derived in this way. The trigram is constructed by plotting a point with D, Q and NA in the $x - y$ plane with cartesian coordinates

$$(x, y) = \left(\frac{Q - NA + 1}{2}, \frac{\sqrt{3}}{2} D\right). \quad (23)$$

This corresponds geometrically to the triangular hyperplane $D + Q + NA = 1$ in the three-dimensional (D, Q, NA) cartesian system, when D, Q and NA are all positive. The three dashed lines in the figure correspond to the lines along which $D = Q$, $D = NA$ and $Q = NA$. So almost axisymmetric dipolar planets, like Saturn, end up close to the point D in the trigram, whereas planets with their magnetic field mainly in the non-axisymmetric components like Uranus and Neptune, which have D and Q small compared to NA, end up close to the point NA.

We first add the solar system planets to our trigram indicated by black stars, by calculating D,Q and NA from the known Gauss coefficients. We adopt the measurements from Cassini (Saturn), Voyager I (Jupiter), Voyager II (Neptune and Uranus) and Messenger (Mercury) (Connerney, 2007; Wicht and Heyner, 2014). The values for the Earth's magnetic field up to degree and order 4 are taken from the 10th IGRF model (Macmillan and Maus, 2005). It can be seen in Fig. 9 that the peculiar field of Saturn has a distinct position closest to the top corner. The Cassini measurements are best modelled with a 3 Gauss coefficient model which are entirely axisymmetric and which have strongly dominant equatorial antisymmetry ($g_{10}$, $g_{30}$). For Mercury only a few axisymmetric Gauss coefficients are used to reconstruct the data, where the comparable strength of equatorial symmetric and antisymmetric field contributions lead to a hemispherical asymmetry of the magnetic field, displacing Mercury from D towards the point Q. For Jupiter the maximal degree and order of the Gauss coefficients is 4 (see e.g. Connerney (2007) for an overview) and for the Earth the field is accurately known. Both have fields dominated by a slightly tilted dipole, but further show significant non-axisymmetric (NA) features. Their small symmetric components mean Q is small, so both Earth and Jupiter are displaced from D in the NA direction. Lastly, the ice giants have rather irregular, non-axisymmetric dynamos positioned in the left bottom corner of the magnetic tri-gram. Note that apart from the Earth, all we have are snapshots of their current position on the trigram.

The values of D, Q and NA from all the models in Table 1 have been added into the trigram Fig. 9. To make it easier to identify the points, the D, Q and NA values, and their corresponding symbols are listed in Table 1. We find the magnetic field solutions of our numerical models fall mainly into two groups. The dipole dominated fields are in the top corner, with small Q, stretching out towards NA. The numerous solutions with dominant quadrupolar symmetry are found at the bottom close to NA-Q line.

For the models with strong dipole contribution, only those with high $r_d$ are found to remain stable over time. Whereas the model 13 closest to that in Jones (2014) (grey circle with black centre in Fig. 9), aligns well with the measured Jupiter field our model with high $r_d$ and stable dipole field (model 6.1, light green pentagon with grey surround) had larger $Pr$ causing larger non-axisymmetric contributions. Indeed, we have not found a single case with intermediate or small $r_d$ harbouring a dominant dipole which is stable in time. However, the more common quadrupolar modes are typically stable in time and do not reverse. The rather peculiar bottom-heated hemispherical wave model (no. 8.1 or type VII in Table 1 and in Fig. 6(f)) has strong non-axisymmetric contributions and lies as it should on the $D = Q$-diagonal (dark red triangle with grey interior).

The preference between dipolar dominated and quadrupolar dominated fields is not well understood, and Fig. 9 shows a rather complex behaviour. The standard set of models (group 2 in tab 1), analysed earlier in Sections 4.2, 4.1 and Fig. 3, is controlled by the Rayleigh number of $Ra = 1 \cdot 10^7$ and Prandtl number of $Pr = 0.25$. The $r_d = \infty$ model has a small-scale dynamo with strong non-axisymmetric contributions (dark green cross close to NA). Now if $r_d$ is reduced to 0.94, the field is more large-scale and the dipole is more dominant. In general if $r_d$ is lowered the models tend towards the axisymmetric solutions because we analyse the field at the surface not at the top of the dynamo region. Hence small-scale modes are systematically weaker, as they decay faster in the potential field. If the conductivity drop-off is further lowered towards $r_d = 0.82, 0.72$, the leading order parity jumps to quadrupolar parity (dark green pentagon and square close to the bottom). Surprisingly, if the magnetic shell is further thinned to $r_d = 0.67$ or 0.62 the solution undergoes another parity inversion and the dynamos are strongly dipolar (dark green diamond and triangle in the top corner). These dipolar fields found for small $r_d$ are not steady dynamos, but are dynamo waves migrating towards the equator. They can be highly axisymmetric with $>90\%$ of their magnetic energy in the $m = 0$ component. We compare our models to observations of Saturn in more detail in the next section.

The reasoning that models with a thicker hydrodynamic shell (smaller $r_d$) appear generally more axisymmetric on the surface due to the drop-off in the non-metallic region assumes that at the surface of the dynamo region the magnetic field morphology is unaffected by the value of $r_d$. This was obviously not true for the models in group 2 with $Ra = 1 \cdot 10^7$. However, we also found surprising behaviour when the driving was enhanced to $Ra = 1.5 \cdot 10^7$ (blue coloured symbols). All of those dynamos are dominantly quadrupolar with irregular or erratic dipolar components. Even though the model with the lowest conductivity drop-off $r_d = 0.62$ has the thickest hydro shell and hence the largest potential field decay, it shows the highest degree of non-axisymmetry. This trend is visible for all models up to $r_d = 0.82$ in that set, where the most axisymmetric fields are found for the thickest dynamo regions (large $r_d$). As an interpretation, we suggest that at this higher $Ra$ there is enhanced flow in the hydrodynamic shell, which leaks into the dynamo region and disrupts the large-scale field more efficiently when the hydrodynamic shell is larger.

Another clear trend visible in Fig. 9 is that increasing the convective vigour while keeping $r_d$ fixed increases the non-axisymmetry of the field. This can be clearly seen for the squares ($r_d = 0.72$, group 5 in Table 1) and pentagons ($r_d = 0.82$, group 6 in Table 1) in the plot. Larger convective vigour leads to more energetic flows, hence larger magnetic Reynolds numbers and more small-scale fields.

We also explored the effect of varying $Pm$. If $Pm$ is reduced much below the value 3 used in the group 2 runs, say down to $Pm = 1$, while $Ra$ is held constant, the enhanced magnetic diffusion leads to dynamo collapse. We can compensate for the enhanced diffusion by increasing $Ra$, so magnetic field generation is supercritical again, but this also increases the non-axisymmetry. Boussinesq models suggest that dipolar solutions can be obtained at lower $Pm$, but only by going to smaller $E$ (Christensen and Aubert, 2006) which is computationally expensive and so not pursued here.

$Rm$ can be tuned also by choosing larger $Pm = 5$ (group 7) instead of 3. Comparing, e.g. model 2.1/2.3/2.5 with the equivalent $Pm = 5$-run (model 7.1/7.2/7.3) shows an enhanced NA-value, where the dominant field parity is conserved. Interestingly, the oscillating dipole wave (2.1) is transformed into a steady octupolar dynamo (7.1).

If the convection is driven from the inner boundary rather than homogeneously inside the shell, the flow is also more energetic mimicking the strong-inertia behaviour (models 2.5 vs 8.4 or 4.3 vs 8.2) with larger contributions to non-axisymmetric magnetic energy. Further, model 8.1 from that group shows the peculiar hemispherical wave dynamo.

Also higher $Ra$ usually leads to more quadrupolar fields. The effect of making $Pr < 0.25$ seems more complex. Whereas lowering $Pr$ for $r_d = 0.62$ and $Ra = 9 \cdot 10^6$ (light green triangles) yields a transition from a dipolar-wave dynamo to a stable quadrupole (group 4 in Table 1), if both $Ra$ and $Pr$ are changed, $Ra = 1 \cdot 10^7 \rightarrow 6 \cdot 10^6$ and $Pr = 0.25 \rightarrow 0.15$, a quadrupole (dark green square, model 2.3) is transformed into a dipolar wave (red square, model 1.2). This suggests higher $Ra$ favours quadrupoles over dipoles for the same $r_d$ and $Pr$, though if $Ra$ is increased too much the dynamo becomes small-scale.

In general the two big clusters of models in the magnetic trigram suggests that models which are fairly axisymmetric are either dominated by dipolar or by quadrupolar symmetry. Mixed models with roughly equal contributions of D and Q are only found in the peculiar hemispherical models. A comparable contribution of both $D$ and $Q$ normally only occurs for small scale very non-axisymmetric dynamos. Note that in our models, the strongest axisymmetry in the models is usually (but not invariably) found for smaller $r_d$, where there is stronger shear and stronger potential field decay of small-scale modes. However, the more axisymmetric models can be either strongly dipolar or strongly quadrupolar.

### 4.5. Saturn-like models

Can our models with small $r_d = 0.62/0.67$ have Saturn-like magnetic fields? The peculiar field of Saturn seems entirely dominated by axisymmetric modes at the surface. The tilt of the dipole is marginal (Cao et al., 2011) if it exists at all. Also, only the first three axisymmetric Gauss coefficients have been measured, and the field is only slightly equatorially asymmetric. These observations suggest an axisymmetric dipole of strength $g_{10} = 21\,\mu T$. The axisymmetric octupole $g_{30} = 2.2\,\mu T$, roughly ten times smaller than the axisymmetric dipole, and unlike the Earth it has the same sign as the dipole, suggesting that the magnetic field is more concentrated near the poles than that given by a purely dipole field (Connerney, 2007). There is an even weaker quadrupole component, $g_{20} = 1.5\,\mu T$. As shown in the magnetic trigram (Fig. 9), mod-els with a small, Saturn-like value of $r_d$ are dominated by equatorially antisymmetric (dipole family) field components. However, they typically periodically reverse with a quarter of the magnetic diffusion time as the typical period. If a steady dipole field is required to model Saturn, then these models without a stably stratified layer cannot do the job. However, our observations of Saturn span only a minute fraction of a magnetic diffusion time. Fig. 10 shows the simulation results for the time evolution of the first three Gauss coefficients at the upper bound of the simulation domain. The input parameters for this particular model are found in Table 1, model 2.1.

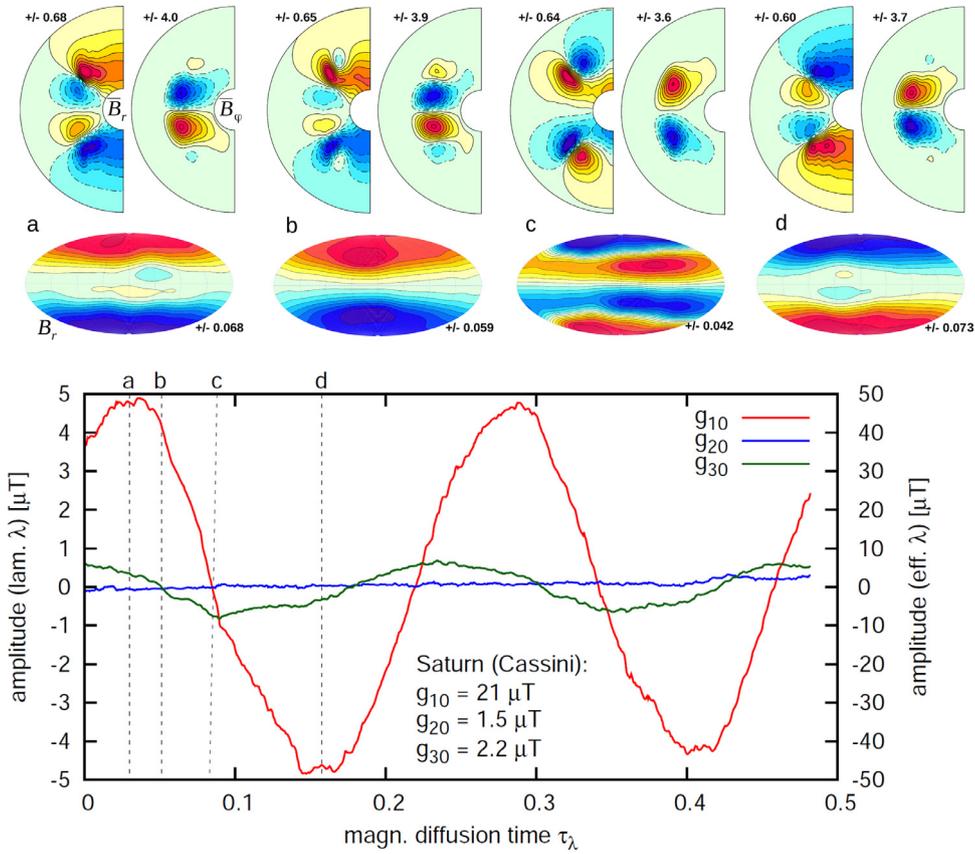

**Fig. 10.** Bottom: Time evolution of the first axisymmetric Gauss coefficients (dipole, quadrupole, octupole for a model from type III). The magnetic field is rescaled using a laminar electrical conductivity (left ordinate) or an effective value (right ordinate) derived from the secular variation. Top: zonally averaged radial and azimuthal field and radial field at the surface for the four snapshots indicated in the time series. The model used is no. 1.2 from Table 1.

To rescale magnetic field amplitude and the time, the simplest way is to use the laminar value of the magnetic diffusivity assuming a metallic electrical conductivity ($\lambda_m = 0.54 \, m^2/s$) (Jones, 2014). The magnetic field strength is rescaled by $\sqrt{\Omega \rho_m \mu_0 \lambda_m}$ and the (magnetic diffusion) time by $\tau_m = D^2/\lambda_m$. The values for Saturn are: $D = 4.698 \cdot 10^7$ m, $\Omega = 2.633 \cdot 10^{-5}$ 1/s and $\rho_m = 919 \, kg/m^3$ (Nettelmann et al., 2013). For laminar $\lambda_m$ the dipole mode reaches then $\pm 5 \, \mu T$, the magnetic diffusion time would be 126 Myrs and the oscillation period is 30 Myr as shown in Fig. 10 using the left ordinate. The relations between the individual Gauss coefficients are correctly reproduced as $g_{10} \approx 10 \cdot g_{30} > g_{20}$, however the amplitude is underestimated by a factor four.

It is likely that small scale turbulence enhances the diffusivity in the metallic region of giant planets hence reduces the magnetic diffusion time substantially and renders the above estimate using laminar $\lambda$ somewhat questionable. As an alternative and more realistic approach, Holme (2007) suggested comparing the observed mean secular variation of the dipole with that of the model to find an 'effective' magnetic diffusivity. Applied to Jupiter by Jones (2014), this method somewhat over-predicted the magnetic field strength, but it is based on more realistic assumptions. Cao et al. (2011) found no clear evidence of secular variation of the Saturn magnetic field, however for comparison we use the suggested end-member maximal value of 2.8 nT/yr. Assuming $g_{10} = 21200$ nT this gives a mean change of the dipole of 0.013% per year. Within the oscillating model we find a root-mean-square $\dot{g}_{10}/g_{10} = 157$ giving an effective magnetic diffusivity $\lambda_e = 60 \, m^2/s$, that is roughly 100 times higher than the laminar value. Expressed in terms of the magnetic field strength (Fig. 10, right ordinate) the dipole oscillates between $\pm 50 \, \mu T$ - a much better estimate. Also the magnetic diffusion time is reduced to 1 Myrs and the oscillation period to 250 kyrs. The models closest to Saturn in terms of the believed conductivity distribution, magnetic field amplitude and symmetries are therefore oscillatory dynamos. The observations of Saturn's magnetic field and hence its time-dependence only span about 30 years, whereas our Saturn model will only show significant secular variation on time-scales of tens of thousand years. With the data available, an oscillatory dynamo wave model is just as satisfactory in matching the observations as a steady dynamo.

However, there is one feature where our models differ from the observed field. The non-axisymmetric components of Saturn's field are now constrained to give a dipole inclination of less than $0.06°$ (Cao et al., 2011), equivalent to $g_{11}$ and $h_{11}$ being less than 20 nT. While the non-axisymmetric components in our Saturn-like model 4.3 are indeed small compared to the axisymmetric components (see Fig. 10), they are not that small. The model dipole tilt is typically around $1-2°$, but there are times when it is some-what less. It therefore appears that although our models without a stable layer are capable of getting the axisymmetric components right, they cannot reduce the non-axisymmetric components to the extremely low values indicated by the recent observations.

Fig. 10 also plots snapshots during the time-evolution of the wave-like dynamo which is dominated by equatorial antisymmetric modes. At snapshot (a), the dipole is maximal and the octupole relates to the inverse flux patches in either hemisphere close to the equator. Those patches disappear at snapshot (b), where the octupole crosses zero. A new pair of inverse flux patches is created at higher latitudes, starting to push the regular flux patch equatorward. At snapshot (c), the dipole is exactly zero, representing equal strength of inverse and regular flux in each hemisphere. Note that the axisymmetric azimuthal field ($\bar{B}_\phi$) is already reversed at this time. The inverse flux patches grow in amplitude and finally reach their maximal strength in snapshot (d), where the dipole is maximal and of inverse sign compared to snapshot (a). Fig. 10 also shows that the radial field at the surface remains predominantly axisymmetric during the evolution. However, there is some weak azimuthal variation, most clearly visible as inverse flux patches at the equator. The field concentration in the higher latitudes due to the dipole and octupole having the same sign is found in our models as well as in the observed Saturn magnetic field.

The observed magnetic field is surprisingly similar to the mean-field models of the solar magnetic field, despite the fact that the Sun is not a rapid rotator, and has much stronger convection than giant planets. We observe a consistent positive kinetic helicity in the dynamo region for models with variable conductivity and found strongly increasing cylindrical shear. Both are proposed as key features of the classic Parker-wave dynamo (Parker, 1955). Even though our models are different in terms of the setup and do not fulfil the mean-field assumptions, such as scale separation, we find some similarities. The shear in our models is purely cylindrical and increases outwards ($\partial \bar{u}_\phi / \partial s > 0$) due to the emerging differential rotation in the hydrodynamic exterior. In the framework of mean field dynamos, the waves migrate equatorward if the poloidal field generation (called the $\alpha$-effect) times the shear is negative (Stix, 1976). Fig. 10 shows snapshots of the azimuthally averaged radial field as a proxy for the poloidal field and zonal field for the toroidal field. The propagation direction is in line with the classic theoretical prediction of equatorwards propagation if we make the common (but not obvious) assumption that $\alpha \propto -h$. Further it was suggested by Yoshimura (1976) that the toroidal field is ahead by a phase shift $\pi/4$ when the shear is simply ($\partial \bar{u}_\phi / \partial s) > 0$. The snapshot (c) in Fig. 10 is taken when the dipole is zero, and the zonal field is already reversed, hence the toroidal field is indeed ahead as predicted.

## 5. Discussion and conclusions

We have performed a suite of numerical simulations for the dynamo generated magnetic fields of rapidly rotating, convecting planets. The models have a range of conductivity distributions, varying from Jupiter-like distributions, which have significant conductivity out to $r_d = 0.94 r_o$, to Saturn-like distributions with a conductivity drop-off at $r_d = 0.62 r_o$. As in Boussinesq models, organised large-scale fields only occur in the rotationally dominated regime (sufficiently low Rossby number). If the Rayleigh number is increased at fixed Ekman number, the Rossby number becomes too large and the generated magnetic field becomes small-scale. Within the large-scale field regime, there is a remarkable diversity of magnetic field configurations. We have extended the classification of the emerging magnetic fields beyond just measuring the relative dipole strength in order to understand these complex dynamos. Butterfly diagrams were used to clarify the equatorial symmetries and time-dependencies, and we introduce a triadic scheme based on the leading magnetic field symmetries (axisymmetry and equatorial symmetry) to distinguish the various solutions ('magnetic trigram', see Fig. 9). Our choice of parameters (e.g. Prandtl, Ekman and Rayleigh number) revealed many different magnetic field solutions. It appears that for a deeper conductivity drop-off (smaller $r_d$), stronger shear emerges within the hydrodynamic exterior and this shear affects the induction process at the interface. This leads to unforeseen alterations of the leading order field parities and temporal evolution of the interior dynamo.

For Jupiter-like models, strong steady dipole dominated fields are possible. As $r_d$ is lowered, there is a whole belt of $r_d$ values between Jupiter-like and Saturn-like values where the magnetic field has mainly quadrupole symmetry. Then as $r_d$ is further reduced to Saturn-like values, oscillatory dipolar fields become preferred, with a period of $O(10^5)$ years. There is therefore a significant range of giant planet masses, between those of Jupiter and Saturn, where quadrupolar magnetic fields predominate. This raises the intriguing possibility that the reason we don't have any planets in our solar system which are strongly quadrupolar is that both Jupiter and Saturn happen to lie just outside the quadrupolar belt. Exoplanets in the appropriate mass range may have quadrupole dominated magnetic fields.

We do not yet fully understand how the morphology of these organised giant planet fields is determined. However, in the course of the investigation we have found a number of significant features which we believe will play a part in the final story. Although we cannot predict the magnitude of the zonal flows in giant planets, our models consistently show that having a deep hydro-region leads to broader, faster equatorial jets, just as Saturn has a broader, faster equatorial jet than Jupiter. The powerful jet arises because in the hydro-shell the Reynolds stress driving the flow is only balanced by the weak (turbulent) viscosity. As we might expect, the pattern of the equatorial jet in our models is not identical to those in real planets, being somewhat too broad, but nevertheless we believe that the strong zonal flow, which penetrates into the edge of the metallic region in lower $r_d$ models, is influencing the dynamo. It is known that strong shear leads to oscillatory dynamos being preferred over steady dynamos, and this may be why our Saturn-like models are always time-dependent. In the deep interior, the Maxwell stresses predominate over the Reynolds stresses, and the differential rotation is reduced there. The balance in the azimuthal momentum equation is then mainly between Maxwell stress and advection of angular momentum (see Fig. 5), and this leads to a meridional circulation which can affect the dynamo.

Another significant result is the helicity reversal found in the metallic regions of our giant planet models (see also Duarte et al., 2016). Our results suggest, that the deep dynamo region, where the Lorentz force relaxes the rotational constraint to some degree, is decoupled from the hydrodynamic outer shell. This is realised when the axial vorticity flips as a convective column extends through the interface. As a consequence, the helicity in the magnetic region is determined by converging/diverging flows near the equatorial plane rather than from the spherical boundaries. This may be also connected with the existence of the Maxwell stresses and the meridional circulation in this region, and it is very likely connected with the preference for quadrupolar over dipolar fields. Sreenivasan and Jones (2011) proposed that a preference for dipolar fields over quadrupolar fields in Boussinesq models was created by the helicity being enhanced by dipolar fields and quenched by quadrupolar fields. In our anelastic models the effect of the magnetic field is so strong it completely reverses the helicity. It is not unreasonable that this could completely alter the relative effects of quadrupolar and dipolar magnetic fields on the helicity, and hence on the dynamo process. It is also possible that the zonal flow leaking into the metallic region from the hydro-region could contribute to the helicity reversal.

We have also found an anelastic hemispherical dynamo, first seen in Boussinesq-models by (Busse, 2002). In our models, the key to the existence of this type of dynamo seems to be the induced meridional circulation. We have not conducted an extensive search for these hemispherical dynamo wave models, but if they can be explained as an interaction between the magnetic field and the meridional circulation (see Fig. 7), it may be possible to shed more light on these slightly exotic dynamos. For the great majority of our models, either the dipole or quadrupole components dominated, though we did find a series of quadrupolar models with an oscillatory dipolar component in combination with the basic steady quadrupole (see Fig. 6).

We had a mixed success in our attempts to find a model which reproduced the observed magnetic field of Saturn. No steady predominantly axisymmetric dipolar solutions were found at $r_d$ values like Saturn's. However, since the oscillating dynamos we found have such a long period compared to our observation window, this is no great problem. It is possible that steady solutions in Saturn models exist at lower Ekman number with some combination of $Pr$ and $Pm$, but the parameters might have to be quite extreme to prevent the strong zonal flow leaking into the dynamo region. Within our oscillatory Saturn models, it is possible to find significant intervals of time where the ratios of the three known axisymmetric Gauss coefficients are Saturn-like. We have therefore succeeded in deriving a plausible model for the axisymmetric components of the field. Where the model fails is in the size of the non-axisymmetric components, which are much too large. Again, it may be possible to reduce the non-axisymmetric components by the factor of 10 needed by going to more extreme parameter values, but this would be very challenging computationally.

There is a difficulty in using the surface zonal flow to constrain giant planet dynamo models. The convective flow speed can be estimated from the heat flux emerging from the surface, giving a typical speed of $10^{-2}$ ms$^{-1}$ (e.g. Christensen and Aubert, 2006). This is also the typical flow speed estimated from the secular change of Jupiter's magnetic field using data from Jovian space mission (Jones, 2014; Ridley and Holme, 2016). If the flow in the metallic region was significantly faster than this, it would have to be aligned with the magnetic field in such a way as to disguise the resulting secular variation, which seems unlikely. Jupiter has surface zonal flow speeds of about $10^2$ ms$^{-1}$, $10^4$ times greater than the convective speeds, and the ratio for Saturn is likely to be even larger than that. It is not possible to reach such large ratios in numerical dynamo simulations, because very small values of Ekman and magnetic Prandtl numbers would be required, and it is then not possible to resolve the resulting small length-scales. However, It is possible to achieve realistic Rossby numbers for the zonal flows, e.g. Heimpel et al. (2005), but only at the expense of increasing the convective heat flux to unrealistically high values to offset the excessive viscous dissipation. Nevertheless, although the magnitude of the zonal flow cannot be reliably estimated using numerical dynamo models, it is still of interest to investigate the pattern of the surface zonal flows generated, as these do relate to the observed pattern.

As a future prospect, we aim for a clearer understanding of the rather complex internal dynamics, especially the magnetic field induction process, the magnetic field parity preference and model parameter dependencies that our simulations have uncovered. As comparable models have successfully reproduced Jupiter's magnetic field (e.g. Jones, 2014), a primary target of research will be to extract the crucial ingredients for a successful Saturn model. From our results, it looks as though a stably stratified layer with differential rotation, along the lines proposed by Stevenson (1982), is still the most natural way to explain Saturn's extraordinarily axisymmetric surface field. There is still no consensus in the high pressure physics community as to whether a stably stratified layer exists in Saturn. For the pressure and temperature ranges appropriate to Saturn, hydrogen and helium become immiscible at a pressure level of 1 Mbar, or 67% of the planet's radius (Stevenson and Salpeter, 1977; Nettelmann et al., 2013) possibly leading to a helium-depleted upper zone and a helium-enriched deeper zone. Lorenzen et al. (2011) found that the transition to immiscibility coincides with the transition to metallicity, at the 67% radius level above which the conductivity drops off rapidly. Whether the emerging helium rain is remixed at a greater depth (Lorenzen et al., 2011), in which case a stable layer is likely, or whether it forms a sediment at the rocky core surface, see e.g. Püstow et al. (2016), remains debated. However, as the alternative scenario of including the effect of stable layers is a realistic possibility, it will be studied further in a subsequent paper.

## Acknowledgments

WD and CAJ gratefully acknowledge insightful discussions with K. Hori, D. W. Hughes, S. M. Tobias, J. Wicht and L. D. V. Duarte. Further we thank the reviewers for their insightful comments and suggestions significantly improving the manuscript. The authors are supported by the Science and Technology Facilities Council (STFC), 'A Consolidated Grant in Astrophysical Fluids' (reference ST/K000853/1). This work was partially undertaken on ARC1, part of the High Performance Computing facilities at the University of Leeds, UK. This work also used the DiRAC Data Centric system at Durham University, operated by the Institute for Computational Cosmology on behalf of the STFC DiRAC HPC Facility (www.dirac.ac.uk). This equipment was funded by a BIS National E-infrastructure capital grant ST/K00042X/1, STFC capital grant ST/K00087X/1, DiRAC Operations grant ST/K003267/1 and Durham University. DiRAC is part of the National E-Infrastructure.